\documentclass[10pt,twocolumn,twoside]{IEEEtran}
\usepackage{amsmath,amsfonts,amssymb,amsbsy,bm,paralist,theorem,color}
\usepackage{graphicx,algorithmic,algorithm,relsize}
\usepackage{stfloats}
\usepackage{multicol}
\usepackage{multirow}
\usepackage{subfigure}
\usepackage{dsfont}
\usepackage{cite}
\newtheorem{lem}{Lemma}
\newtheorem{rem}{Remark}
\newtheorem{thm}{Theorem}
\newtheorem{cor}{Corollary}
\newtheorem{prop}{Proposition}

\newcommand{\SINR}{{\rm{SINR}}}
\newcounter{TempEqCnt}
\DeclareMathOperator*{\st}{s.t.}

\graphicspath{{fig/}}

\definecolor{orange}{RGB}{255,107,0}
\definecolor{green}{RGB}{0,160,20}

\begin{document}
\title{Transmission Energy Minimization for Heterogeneous Low-Latency NOMA Downlink}
\date\today
\author{
	Yanqing Xu, Chao Shen, \IEEEmembership{Member,~IEEE,} Tsung-Hui Chang, \IEEEmembership{Senior Member,~IEEE,} Shih-Chun Lin, \IEEEmembership{Senior Member,~IEEE,} Yajun Zhao, and Gang Zhu\\

\thanks{This work has been accepted by IEEE Transactions on Wireless Communications, Otc. 2019.
	
	Part of this work has been presented in IEEE Global Communication Conference (GlobeCom) Workshop on Ultra-Reliable and Low-Latency Communications, Dec., 2017, Singapore. \cite{GlobeCom2017-XU}. 
	The work of T.-H. Chang was supported in part by the NSFC, China, under Grant 61571385 and Grant 61731018, and in part by the Shenzhen Fundamental Research Fund under Grant No. ZDSYS201707251409055 and No. KQTD2015033114415450.
	The work of S.-C. Lin was supported in part by the Ministry of Science and Technology (MOST), Taiwan, under Grant 107-2628-E-011-003-MY3.

Yanqing Xu, Chao Shen and Gang Zhu are with the State Key Laboratory of Rail Traffic Control and Safety, Beijing Jiaotong University, Beijing, China (email: \{xuyanqing, chaoshen, gzhu\}@bjtu.edu.cn).

Tsung-Hui Chang is with the School of Science and Engineering, The Chinese University of Hong Kong, Shenzhen, China,  and also with the Shenzhen Research Institute of Big Data, Shenzhen, China (email: tsunghui.chang@ieee.org).

Shih-Chun Lin is with the Department of Department of Electronic and Computer Engineering, National Taiwan University of Science and Technology, Taipei, Taiwan (email: sclin@ntust.edu.tw).

Yajun Zhao is with the Algorithm Department, Wireless Product R\&D Institute, ZTE Corporation, Shenzhen, China (email: zhao.yajun1@zte.com.cn).
}}

\maketitle

\begin{abstract}
This paper investigates the transmission energy minimization problem for the two-user downlink with 
strictly heterogeneous latency constraints. To cope with the latency constraints and to explicitly specify the trade-off between blocklength (latency) and reliability the normal approximation of the capacity of finite blocklength codes (FBCs) is adopted, in contrast to the classical Shannon capacity formula. We first consider the non-orthogonal multiple access (NOMA) based transmission scheme. 
However, due to heterogeneous latency constraints and channel conditions at receivers, the conventional successive interference cancellation may be infeasible. 
We thus study the problem by considering heterogeneous receiver conditions under different interference mitigation schemes and solve the corresponding NOMA design problems.
  It is shown that, though the energy function is not convex and does not have closed form expression, the studied NOMA problems can be globally solved semi-analytically and with low complexity.
Moreover, we propose a hybrid transmission scheme
that combines the time division multiple access (TDMA) and NOMA. Specifically, the hybrid scheme can judiciously perform bit and time allocation and take TDMA and NOMA as two special instances.  
To handle the more challenging hybrid design problem, 
we propose a concave approximation of the FBC rate/capacity formula, by which we obtain computationally
efficient and high-quality solutions.  Simulation results show that the hybrid scheme can achieve considerable transmission energy saving compared with both pure NOMA and TDMA schemes.
\end{abstract}


\section{Introduction}
The ultra-reliable and low-latency communication (URLLC) is one of
the emerging application scenarios in 5G
\cite{3GPP-5G}\cite{Popviski-2016-Procceding}, where the system
promises to serve multiple autonomous machines with high
reliability and low
latency\cite{popovski-urc-2014,popviski-urllc-network,she-magazine-2017}.
The traffic of such an URLLC system is drastically different from
that of the human-centric 4G LTE. More specifically, the
communication is required to have no less than $99.999\%$
reliability (that is, $10^{-5}$ packet error probability), no
longer than $1$ms latency, and small packet size (such as $32$
bytes) \cite{3GPP-URLLC1}. Therefore, especially for multi-user
channels, new system architectures and transmission schemes
compared to the traditional human-centric communications are
required to achieve the URLLC specifications. Moreover,
energy-efficiency is a key performance indicator of 5G
communications \cite{Andrews-2014}, and it is important to develop
new energy-efficient transmissions for URLLC multi-user channels.

However, the existing energy-efficient transmission protocols only
target at human-centric communications, and are based on the
traditional Shannon capacity formula, such as
\cite{Wang-2013}\cite{Chen-2009}. The Shannon capacity is accurate
only when the codeword blocklength is infinitely long
\cite{Shannon-2001}\cite{Book_Cover}, and thus not applicable to
the URLLC systems. Therefore, it is well motivated to investigate the
system design using a finite blocklength code (FBC). Recently,
a tight lower bound of the maximal achievable rate of a FBC in the Gaussian channel has
been characterized in \cite{2010-Polyasiki-TIT}, which is named as ``normal approximation'', and it is later
extended to the ergodic fading channel \cite{ISIT11-Polyanskiy}
and the outage-constrained slow fading channel
\cite{yury-FBC-fading}. In this work, the normal approximation of the FBC rate/capacity formula is used. 
The new capacity formula for FBC
\cite{2010-Polyasiki-TIT} explicitly characterizes the
relationship between transmission rate, codeword blocklength and
decoding reliability, and thus is particularly suitable for
evaluating the performance of the URLLC system. Moreover, 
the information-theoretic capacity result in
\cite{2010-Polyasiki-TIT} can be practically approached via the
polar code with short blocklength \cite{polar-tse}\cite{polar2}. The normal approximation of FBC has been
successfully applied to the study of various communication
scenarios with strict latency constraints, as in
\cite{Ostman-2019,Xu-2016,Ozcan-2013,Gursoy-2013,Makki-2015,Makki-tcomm,Hu-TWC-2016,Hu-TVT-2016,she-twc-2017}.
In the context of energy-efficiency, \cite{Xu-2016} considered the
energy-efficient packet scheduling problem in a point-to-point
system and showed that using the classical Shannon capacity
\cite{Shannon-2001} can significantly underestimate the energy
with the FBC.

As a promising enabling technique of 5G, the
non-orthogonal multiple access (NOMA), which allows multiple users
to transmit simultaneously over non-orthogonal channels, has been extensively studied
\cite{Ding-Magazine}\cite{XU-TSP-2017}. Indeed, for the downlink
channel, the superposition coding based NOMA is shown to be
capacity achieving when the blocklength is long \cite{Book_Cover}.
Moreover, similar transmission scheme, known as the multiuser
superposition transmission (MUST), has already been approved by
the 3rd generation partnership project (3GPP) \cite{3GPP-MUST}.
Compared to the orthogonal multiple access (OMA), NOMA can exploit
the channel diversity more efficiently via smart interference
management techniques such as the successive interference
cancellation (SIC) \cite{Book_Cover}\cite{Ding-Magazine}. {For uplink system with heterogeneous user latency requirements, NOMA were considered in \cite{ding-mec-1,ding-mec-2}. However, aforementioned NOMA works \cite{Ding-Magazine}\cite{XU-TSP-2017}\cite{ding-mec-1}\cite{ding-mec-2} were based on traditional Shannon capacity formula.
Aiming at URLLC applications,  NOMA system designs with FBC attract lots of attentions
\cite{hu-pimrc-2017,noma-fbc-cl,Sun-2017,Popovski-tcomm}.} In the
downlink channels, \cite{noma-fbc-cl} aims to minimize the common
blocklengths of users while guaranteeing different reliability
requirements; also under equal blocklength constraints, \cite{Sun-2017}
considers  maximization of the effective throughput. Finally,
assuming all users experience equal channel conditions in the
downlink, the FBC transmission by grouping users at the
transmitter and decoding all user messages at each receiver is
considered in \cite{Popovski-tcomm}. All the former works
\cite{hu-pimrc-2017,noma-fbc-cl,Sun-2017,Popovski-tcomm} assume
certain kinds of homogeneity such as the same blocklengths (latency
constraints) or the same channel conditions. In practice, downlink
users may ask for \textit{heterogeneous} quality of service (QoS) \cite{popviski-urllc-network,3GPP-URLLC1,Andrews-2014},
and designing corresponding transmission protocols is crucial.

In this paper, we consider energy-efficient resource allocation
for a two-user heterogeneous NOMA downlink with an FBC. In particular,
based on the superposition coding, we aim to solve the energy
minimization problems subject to heterogeneous latency and
reliability constraints at downlink users. Due to heterogeneous
latency constraints (blocklength) and channel conditions, unlike
\cite{noma-fbc-cl,Sun-2017,Popovski-tcomm}, SIC may not always be
feasible since there exist situations where none of the receivers can perform SIC and decode messages of the other users. 
In view of this, 
we first propose several
achievable interference cancellation management schemes according to whether SIC is
feasible or not. 
While solving the FBC formulated design problem is challenging, we show that the problems have semi-analytical solutions. 
It turns out that the proposed NOMA schemes under heterogeneous latency constraints may be less energy-efficient than the OMA scheme such as  time division multiple access (TDMA), in contrast to \cite{Book_Cover}\cite{noma-fbc-cl}\cite{Sun-2017} where the users have a common latency. To overcome this issue, we present a hybrid
transmission scheme which includes both NOMA and TDMA as special
cases. 
The main contributions of this paper are summarized as follows.

\begin{itemize}
	\item We globally solve the energy minimization problems in (super-position coding based) NOMA downlinks under heterogeneous latency constraints and channel conditions. Though the target energy is a \textit{non-convex implicit function}, the \textit{optimal} blocklengths and powers for users to minimize the transmission energy can still be obtained in semi-closed forms with a low complexity. The key is identifying the monotonicity of the energy function with respect to the code blocklengths with the aid of the implicit function theorem
	\cite{Krantz_Parks02}. Moreover, unlike the solver for TDMA \cite{Xu-2016}, the \textit{feasibility} of our solver for NOMA downlink can be simply checked.
	\item  We propose a hybrid transmission scheme which consists of
	the NOMA and TDMA as special cases, and can be \textit{strictly} better than both.
	However, the corresponding energy minimization is harder than
	that for only NOMA. We then find a \textit{tight concave approximation}
	of the normal approximation of FBC capacity formula, with given blocklength and error probability, and develop a suboptimal but computationally efficient algorithm. 
The developed algorithm has a much smaller complexity
	compared with the one based on naive linear search.
\end{itemize}
\noindent Our simulation results with URLLC settings in 3GPP \cite{3GPP-URLLC1}\cite{3GPP-URLLC2} show that the superposition-coding based NOMA is more energy efficient than the
TDMA when the two users have similar and homogeneous latency constraints.
However, the hybrid scheme
can enjoy the benefits from the both.
{{Finally, similar to observations for TDMA in  \cite{Xu-2016}, using traditional Shannon capacity formula would significantly underestimate the transmission energy in NOMA URLLC downlink.}} Compared to the conference version of this work \cite{GlobeCom2017-XU}, more details for the low-complexity pure NOMA algorithms are provided including the feasibility test presented in Remarks 1 and 2. Moreover, to further decrease the transmission energy of these algorithms, {\textit{new}} algorithms for the hybrid transmission schemes are proposed in Section \ref{sec_hybrid} and their superior energy-savings are demonstrated by new simulations in section IV.

\begin{table}[t]
	\centering 
	\caption{Summary of the notations}
	\label{table1}  
	\begin{tabular}{|c|p{6.6cm}|}  
		\hline 
		Symbols &Descriptions\\  
		\hline
		$N_k$& Number of information bits of receiver $k$ \\
		$m_k$& Code blocklength of receiver $k$ \\	
		$x_k$ & Unit-power coded symbol of receiver $k$\\
		$p_k$ & Transmit power of receiver $k$\\
		$P_{\max}$ & Maximum transmit power of the BS \\
		$\tilde{h}_k$ & Channel coefficient of receiver $k$\\
		$n_k$ & Additive Gaussian noise at receiver $k$\\
		$\sigma_k^2$ & Noise power at receiver $k$\\
		$D_k$& Latency of receiver $k$ \\
		$\gamma_k$ & Received SNR/SINR at receiver $k$\\
		$\epsilon_k$ & Predefined block error probability for receiver $k$\\
		$\hat{m}$ & Minimum blocklength for the normal approximation  of FBC capacity formula holding true\\
		$\bar{\epsilon}_k$ & Overall decoding error probability of receiver $k$\\
		$\Gamma_k (m_k)$ &Continuously differentiable implicit SNR/SINR functions with respect to blocklength of receiver $k$ \\
		$\Gamma_k^{-1} (\gamma_k)$ & Inverse function of the SNR/SINR functions, denoting blocklength \\		$N_{21}$, $N_{22}$ & Number of bits of the two split packets of receiver 2\\
		$m_{21}$, $m_{22}$ & Blocklengths of the two split packets of receiver 2\\
		$x_{21}$, $x_{22}$ & Unit-power coded symbol of the two split packets of receiver 2\\
		$p_{21}$, $p_{22}$ &Transmit power of the two split packets of receiver 2\\
		$\gamma_{21}$, $\gamma_{22}$  &Received SNR/SINR for decoding the two split packets of receiver 2\\
		$\epsilon_{21}$, $\epsilon_{22}$ &Predefined block error probabilities of the two split packets of receiver 2\\
		\hline
	\end{tabular}
\end{table}

{\bf Synopsis:} Section \ref{sec_NOMA} presents the energy
minimization problems of NOMA schemes under heterogeneous user
requirements. The hybrid transmission scheme is investigated in
Section \ref{sec_hybrid} with the convex approximation of the normal approximation of FBC
capacity presented in Section \ref{sec_covexapprox}. Simulation
results with URLLC settings and Conclusions are presented in Section \ref{sec_simu}
and \ref{sec_con} respectively.

For convenience, the involved symbols and the corresponding descriptions are summarized in Table \ref{table1}.

\section{System Model and Energy Efficient NOMA Schemes} \label{sec_NOMA}
\subsection{System model} 
\begin{figure}[!t]
	\centering
	\includegraphics[width=0.7\linewidth]{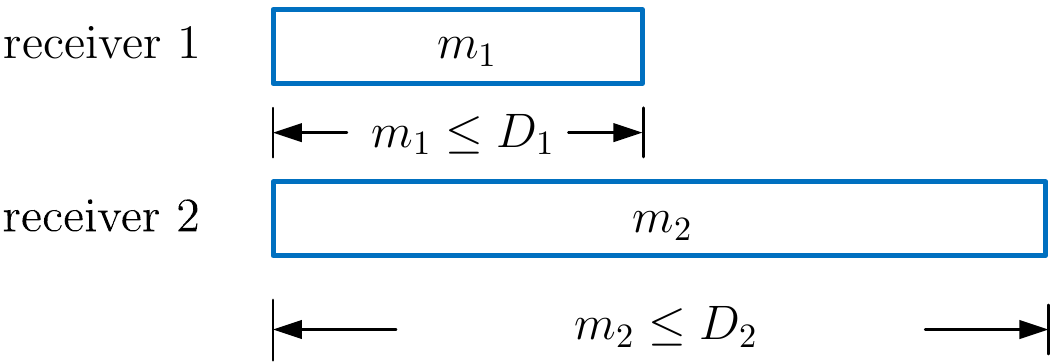}\\
	\caption{ Latency-constrained NOMA downlink, where the deadline $D_2$ of receiver 2 is longer than that of receiver 1. } \label{noma_large_D2}
\end{figure}

We investigate an energy-efficient packet transmission problem in a downlink single-antenna system where a transmitter wants to send two individual messages to two
 receivers respectively, as in Figure \ref{noma_large_D2}. 
 The two receivers are heterogeneous in the sense that they have different transmission latency constraints and different channel gains.
According to the NOMA principle,
the transmitter encodes the $N_k$ message bits for receiver $k$ into a codeword with block length $m_k$ (symbols), $k=1,2$; and transmits the superposition of these two codewords to the receivers. The transmitted signal is then
		$
		\sqrt{p_1} x_{1,i} + \sqrt{p_2} x_{2,i}.
		$
		Here $p_1$ and $p_2$ are the transmission powers allocated to user 1 and 2 respectively, and $x_{1,i}$ and $x_{2,i}$ are the unit-average-power coded symbols at time index $i$ for user 1 and 2 respectively. For simplicity and follow the convention of \cite{Book_Cover}, we remove the time index of the symbols. The received signal for receiver $k$ is given by
\begin{align} \label{eq_downlink}
y_k = \tilde{h}_k (\sqrt{p_1} x_{1} + \sqrt{p_2} x_{2}) + n_k, \quad k = 1,2,
\end{align}
where $\tilde{h}_k \in \mathbb{C}$ is the channel coefficient of receiver $k$ and both the phase and amplitude of $\tilde{h}_k$ is assumed to be perfectly known at receiver $k$, and $n_k \sim \mathcal{CN}(0,\sigma_k^2)$ is the additive Gaussian noise at receiver $k$.
 Different from the traditional downlink schemes \cite{Book_Cover}, 
 due to the \textit{strict} latency constraints imposed on the two receivers,
 the codeword block length $m_k$ must be smaller than $D_k$ symbols (channel uses), $k=1,2$.
 To cope with the new latency constraints, we adopt the normal approximation of FBC capacity formula in \cite{2010-Polyasiki-TIT} since the classical Shannon capacity formula is no longer appropriate.

Besides the encoder, the conventional SIC based decoders in \cite{Book_Cover} also need to be re-designed due to the latency constraints. Note that 
for the two heterogeneous receivers, 
unlike \cite{Popovski-tcomm} the channel gains can be unequal $|\tilde{h}_1| \neq |\tilde{h}_2|$, and unlike \cite{noma-fbc-cl,Sun-2017} the latency constraints can also be different $D_1 \neq D_2$. Thus unlike \cite{Book_Cover}, when $D_1 < D_2$ and $h_1>h_2$, where $h_1 = {|\tilde{h}_1|^2}/{\sigma_1^2}$ and  $h_2 = {|\tilde{h}_2|^2}/{\sigma_2^2}$ are the normalized channel gains at receiver 1 and receiver 2, respectively,
receiver 1 \textit{may not} be able to decode receiver 2's message and cancel the corresponding interference by SIC. Also, the signal $y_2$ received at receiver 2 may not be a degraded (always worse) version of $y_1$. Thus one needs to design decoding strategies according to not only the channel gains $h_k$s but also the heterogeneous latency constraints $D_k$s. Without loss of generality, we assume $D_1 < D_2$ as in Figure \ref{noma_large_D2} and consider two cases in this paper, that is, $h_1 \leq h_2$ and $h_1 > h_2$. The proposed energy efficient transmission schemes are detailed in the next subsection.

\subsection{NOMA Transmission Under Different Channel Conditions} \label{subsec_largerh2}
\noindent \textbf{Case I} ($h_1 \leq h_2$) : Let us start from the case of $h_1 \leq h_2$ 
and receiver 2 performs SIC. Note that the message of receiver 1 is encoded over all $m_1$ symbols and thus one needs to collect all $m_1$ message-carrying symbols for successful decoding from \cite{Book_Cover}.
Since $D_1 < D_2$, receiver 2 can apply SIC to remove $x_1$ if $m_1 \leq m_2$, whereas receiver 1 can only treat interference $x_2$ as noise. Specifically, for receiver 1, interference symbol $x_2$ is modeled as the Gaussian noise \cite[Eq. (198)]{2010-Polyasiki-TIT} in \eqref{eq_downlink}. Thus the achievable rate of receiver 1 under FBC is given by \cite{2010-Polyasiki-TIT}\cite{Xu-2016}
\begin{align} \label{eq:rate_receiver1}
\frac{N_1}{m_1} = \log_2(1 \!+\! \gamma_1) \!-\! \sqrt{\frac{1}{m_1}\left(1 \!-\! \frac{1}{(\gamma_1 \!+\! 1)^2}\right)} \frac{Q^{-1}(\epsilon_1)}{\ln2}, 
\end{align}
where $N_1$ denotes the number of information bits of user 1, $\gamma_1 = \frac{p_1 h_1}{p_2 h_1 + 1}$ is the received signal-to-interference-plus-noise ratio (SINR) for receiver 1, $\epsilon_1$ is the predefined block error probability for receiver 1, and $Q^{-1}(\cdot)$ is the inverse of the Gaussian Q-function. Here we clarify that \eqref{eq:rate_receiver1} is derived from full interference assumption where the whole symbols in the transmission block are with interference. Then the original model in \cite{2010-Polyasiki-TIT} can be applied by changing the SNR with SINR as $\frac{p_1h_1}{p_2h_1+1}$\footnote{ Compared with the AWGN capacity upper-bound in \cite[equation (612)]{2010-Polyasiki-TIT}, achievable rate in \eqref{eq:rate_receiver1} has loss within $\frac{\log(m_1)+\mathcal{O}(1)}{m_1}$}. 

By the principle of SIC, receiver 2 would decode receiver 1's codeword with SINR $\frac{p_1 h_2}{p_2 h_2 + 1}$ in the first stage. Since $h_1 \leq h_2$, the SINR value $\frac{p_1 h_2}{p_2 h_2 + 1}$ is higher than $\gamma_1$ and therefore the error probability of SIC is no larger than $\epsilon_1$.
. By successfully subtracting $x_1$ from $y_2$ in \eqref{eq_downlink} 
receiver 2 then decodes its private message, with probability $1-\epsilon_{2}$, and achieves a rate satisfying
\begin{align}
\frac{N_2}{m_2} = \log_2(1 \!+\! \gamma_2) \!-\! \sqrt{\frac{1}{m_2}\left(1 \!-\! \frac{1}{(\gamma_2 \!+\! 1)^2}\right)} \frac{Q^{-1}(\epsilon_2)}{\ln2}, \label{5}
\end{align}
where $\gamma_2 = p_2 h_2$ and  $\epsilon_2$ is the error probability conditioned on correct SIC. Here we assume that the decoding of user 2's information bits will be erroneous if the SIC fails to decode the interference from user 1 first.
Note that correct SIC needs that the decoding of interference, or user 1's codeword, be successful at receiver 2. 
So the overall decoding error probability of receiver 2, i.e., $\bar{\epsilon}_2$, is upper-bounded by 
\begin{align}
	\bar{\epsilon}_2 &\le  (1-\epsilon_1)\epsilon_2 + \epsilon_1
	= \epsilon_1 + \epsilon_2 - \epsilon_1\epsilon_2. \label{eq_user2errSmallh1}
\end{align}

Based on the above models, the latency-constrained energy minimization design problem{\footnote{Even without SIC, joint minimization of the overall energy consumption of both transmitter and receiver is complicated \cite{cui-2004}. To make optimization problem tractable, we only focus on the energy consumption of the transmitter, as in \cite{Wang-2013,Chen-2009,Xu-2016}.}} for the case of $h_1<h_2$ is formulated as
\begin{subequations}\label{P-noma-case1}
	\begin{align}
	\min_{\{m_k,p_k,\gamma_k\}_{k=1,2}} \quad & m_1 p_1 + m_2 p_2 \label{P-noma-case1 EF}  \\
	\st \quad\quad\quad & F_k(m_k,\gamma_k) = 0,  k = 1,2, \label{p1.1}\\
	& \hat{m} \le m_k, k = 1,2,\label{p1.2}\\
	& m_1 \leq m_2, ~ m_k \le D_k,  k = 1,2,  \label{p1.3}\\
	& p_1 + p_2 \le P_{\rm max}, ~0 \le p_k, k = 1,2,\label{p1.4}\\
	& \gamma_1 = \frac{p_1 h_1}{p_2 h_1 + 1}, ~ \gamma_2 = p_2 h_2, \label{p1.7}
	\end{align}
\end{subequations}
where \eqref{p1.3} are the latency constraints, and \eqref{p1.1} are the normal approximation of FBC capacity constraints with
\begin{align}
F_k(m_k,\gamma_k) & \triangleq  \nonumber\\
&\!\!\!\!\!\!\!\!\!\!\!\!\!\!\!\!\!\!\!\!\!\!\sqrt{\frac{1}{m_k}\!\left(1 \!-\! \frac{1}{(\gamma_k \!+\! 1)^2}\right)}\! \frac{Q^{-1}(\epsilon_k)}{\ln2} \!-\! \log_2(1 + \gamma_k) \!+\! \frac{N_k}{m_k} \label{transformation_1}.
\end{align}
Note that \eqref{p1.1} with $k=1$ corresponds to \eqref{eq:rate_receiver1}, and  \eqref{p1.1} with $k=2$ corresponds to \eqref{5}. Constraints \eqref{p1.2} represents the minimum blocklength constraint for \eqref{p1.1} holding true \cite{2010-Polyasiki-TIT}\cite{Xu-2016} (typically $\hat{m}=100$), while \eqref{p1.4} is the transmission power constraint. 

We should remark that problem (5) is a conservative formulation in the sense that $p_1$ in fact is $0$ and $p_2$ can be $P_{\max}$ after transmitting $m_1$ symbols, but \eqref{p1.4} limits $p_2 \leq P_{\max}-p_1$ for the entire $m_2$ symbols. We will attempt to resolve this issue in Section III by proposing a more sophisticated hybrid transmission scheme.
	Here we will first focus on this pure NOMA scheme to obtain some interesting insights on the NOMA based transmission and solve the corresponding problems with low-complexity (upcoming) Algorithm \ref{alg:bisection} which only needs non-exhaustive bisection search.

Solving problem \eqref{P-noma-case1} is challenging. In particular, the variables are coupled in the constraints in a non-convex and complex fashion. However, in the upcoming Section \ref{subSec_solution1}, we will show how a globally optimal solution to \eqref{P-noma-case1}  can be obtained.

\noindent \textbf{Case II} ($h_1 > h_2$) : As aforementioned, unlike the case of $h_1 \leq h_2$, SIC may not be always feasible when $h_1 > h_2$ and $D_1 < D_2$. Thus we consider two scheduling policies as follows.\\
\noindent \textbf{B.1 Full blocklength for receiver 2:} In this case, we allow $m_1\leq m_2$ since $D_1 < D_2$. Therefore, receiver 1 is not able to perform SIC, but can only treats $x_2$ as noise. Then the energy minimization problem is formulated as 
\begin{subequations}\label{P-noma-case21}
	\begin{align}
	\min_{\{m_k,p_k,\gamma_k\}} \quad & m_1 p_1 + m_2 p_2\\
	\st \quad\quad & \eqref{p1.1},\eqref{p1.2},\eqref{p1.4}, \notag \\
	& m_k \le D_k,  k = 1,2,\label{p3.5}\\
	& \gamma_1 = \frac{p_1 h_1}{p_2 h_1 + 1}, \\
	& \gamma_2 = \frac{p_2 h_2}{p_1 h_2 + 1}.\label{p3.7}
	\end{align}
\end{subequations}

In this case, the first $m_1$ coded symbols for user 2 are superposed with those for user 1 and then transmitted. While for the rest $m_2-m_1>0$ symbols for user 2, they are directly transmitted without those for user 1, that is, the power allocated to user 1 in this period is zero.

\noindent \textbf{B.2 Short blocklength for receiver 2:} In this case, we force 
\begin{align}
m_2 \le m_1. 
\end{align}
Note that $m_1\leq D_1<D_2$, thus the original latency constraint $m_2 \leq D_2$ is automatically satisfied. Under the setting of $m_2 \le m_1$, SIC can be performed at receiver 1 to completely remove the interference from receiver 2. Similar to receiver $1$ in Case I, the overall decoding error probability of receiver 1 is given by
\begin{subequations}
	\begin{align}
		\bar{\epsilon}_1 &\le  (1-\epsilon_2)\epsilon_1 + \epsilon_2\\
	&= \epsilon_1 + \epsilon_2 - \epsilon_1\epsilon_2. \label{eq_user1errSmallh2}
	\end{align}
\end{subequations}
Then the energy minimization problem is formulated as
\begin{subequations}\label{P-noma-case22}
	\begin{align}
	\min_{\{m_k,p_k,\gamma_k\}} \quad & m_1 p_1 + m_2 p_2\\
	\st \quad \quad&\eqref{p1.1},\eqref{p1.2},\eqref{p1.4}, \notag \\
	& m_1 \le D_1, ~m_2 \le m_1,\\
	& \gamma_1 = p_1 h_1, \\
	& \gamma_2 = \frac{p_2 h_2}{p_1 h_2 + 1}.\label{p4.7}
	\end{align}
\end{subequations}

\begin{figure}[!tp]
	\centering
	\includegraphics[width=1.0\linewidth]{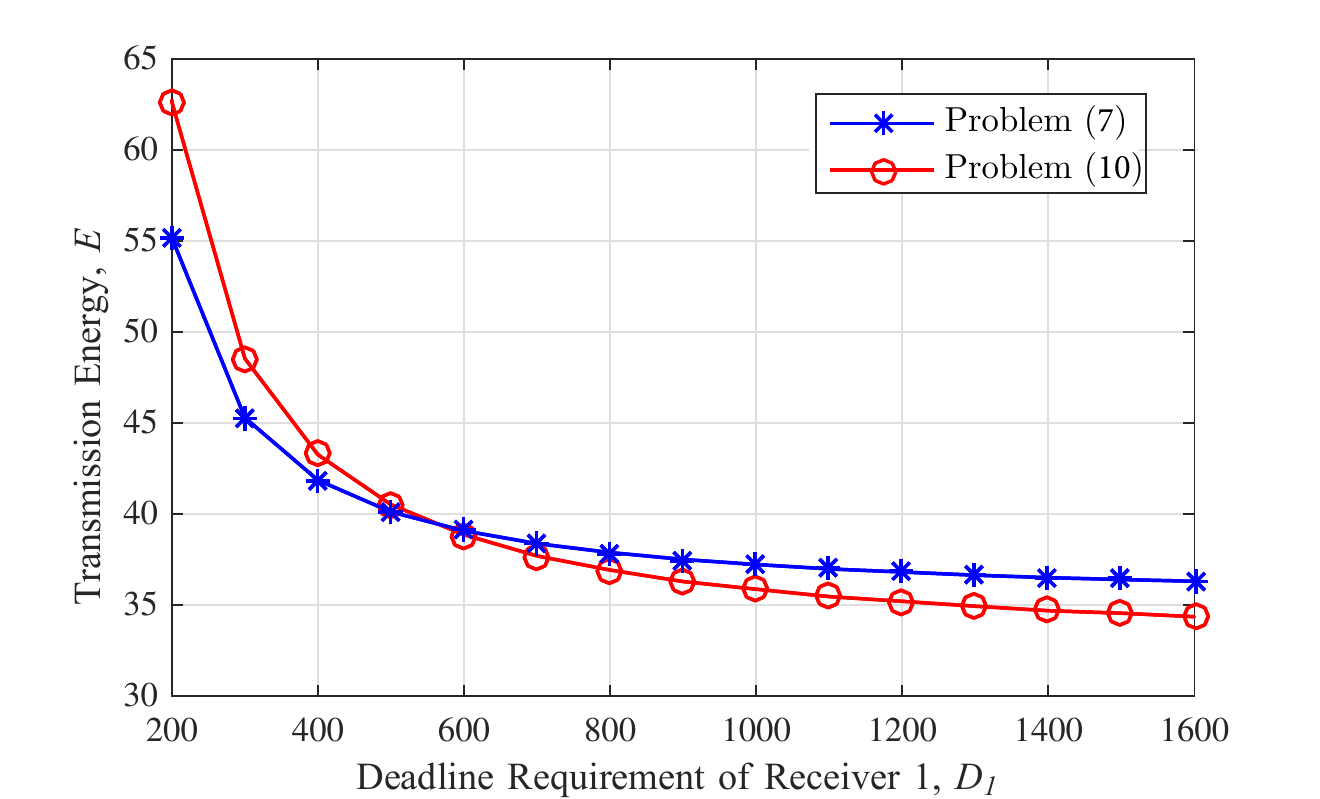}\\
	\caption{Energy consumption comparison of problem \eqref{P-noma-case21} and \eqref{P-noma-case22} with $D_2 = 3800$, $h_1 = 100$, $h_2 = 10$, $\epsilon_1 = \epsilon_2 = 10^{-6}$, $N_1 = N_2 = 256$ bits and $P_{\max} = 40$ dBm.} \label{fig:sic_nosic_noma_case2}
\end{figure}

The solutions of aforementioned two problems are given in Section \ref{subSec_solution2}. 
As will be seen shortly, it turns out that
problem \eqref{P-noma-case22} can yield a smaller transmission energy than \eqref{P-noma-case21} when the two deadlines $D_2$ and $D_1$ are close, thanks to the performance gain brought by SIC. However, when $D_2$ is significantly larger than $D_1$, formulation \eqref{P-noma-case21} can become more energy efficient by benefiting from long code transmission as shown in Fig. \ref{fig:sic_nosic_noma_case2}.

\subsection{Optimal Solutions of NOMA Transmission Problems} \label{Sec_solution}
In this subsection, we present the solutions to the NOMA problems in \eqref{P-noma-case1}, \eqref{P-noma-case21}, and \eqref{P-noma-case22}.

\subsubsection{ Optimal Solutions for Problem \eqref{P-noma-case1}} \label{subSec_solution1}

First, let us briefly recall the implicit function theorem \cite{Krantz_Parks02}  as below 
\begin{thm}\label{thm:IFT}
		Suppose that $f(x,y)$ is a continuously differentiable function with $x \in A$ and $y \in B$ where $A \subseteq \mathcal{R}$ and $B\subseteq \mathcal{R}$ are the domains of function $f$ with $\mathcal{R}$ denoting the Euclidean space.  If for any $x \in A$ there exists a unique $g(x) \in B$ such that $f(x,g(x)) = 0$, then $g(x)$ is differentiable.
\end{thm}
Thus according to Theorem \ref{thm:IFT} and \eqref{p1.1}, there exist continuously differentiable implicit functions $\Gamma_k(\cdot)$ such that
\begin{equation}\label{eq_Gamma_P_MOMA}
\Gamma_k (m_k) = \gamma_k,  k=1,2.
\end{equation}
Note that $\Gamma_k(m_k)$ can be treated as the SINR function with respect to blocklength $m_k$. Thus from \eqref{p1.7}, we have
\begin{subequations}\label{eq:p_gamma}
	\begin{align}
	p_1 &= \frac{\gamma_1 \gamma_2}{h_2} + \frac{\gamma_1}{h_1} = \frac{\Gamma_1(m_1) \Gamma_2(m_2)}{h_2} + \frac{\Gamma_1(m_1)}{h_1},\\
	p_2 &=  \frac{\gamma_2}{h_2} = \frac{\Gamma_2(m_2)}{h_2}.
	\end{align}
\end{subequations}
Then by \eqref{eq_Gamma_P_MOMA}, we can rewrite the target energy of \eqref{P-noma-case1 EF} as a function of block length $m_k$s as
\begin{equation}\label{eq_P_MOMA_target}
\frac{m_1 \Gamma_1(m_1) \left(\Gamma_2(m_2)h_1/h_2+1\right)}{h_1} + \frac{m_2 \Gamma_2(m_2)}{h_2}.
\end{equation}

Now we have the following Lemma.
\begin{lem}\label{lem:energy_monotonicity}
	Function $E_k(m_k) \triangleq  m_k\Gamma_k(m_k)$ is strictly decreasing with blocklength $m_k \in [\hat{m},\infty)$ provided that the error probability $\epsilon_k$ and packet size $N_k$ satisfies
	\begin{align} \label{monotonicity_condition}
	\frac{Q^{-1}(\epsilon_k)}{\sqrt{N_k}} \le \frac{2\sqrt{\ln2}}{4-\sqrt{2}}= 0.64394\cdots.
	\end{align}
\end{lem}
\begin{IEEEproof}
	The proof is relegated to Appendix \ref{app1}.
\end{IEEEproof}
It is worthwhile to note that compared to  \cite[Proposition 1]{Xu-2016}, condition \eqref{monotonicity_condition} is less restrictive as it allows the monotonicity to hold under much milder conditions (e.g., $\epsilon_k \ge 10^{-10}$ and $N_k \ge 100$). Indeed, \eqref{monotonicity_condition} is satisfied in the URLLC system, where the typically required codeword error probability is $10^{-6}$ and the packet size is around $256$ bits ($32$ bytes) \cite{3GPP-URLLC1}.
Based on the monotonicity presented in Lemma \ref{lem:energy_monotonicity}, we can globally solve our problem \eqref{P-noma-case1} as stated in Theorem \ref{thm:noma_case1}\footnote{The ``global solution'' here means that we find the required minimum energy for the proposed NOMA scheme in section II. We do not claim that this is the minimum energy consumption of all possible transmission schemes satisfying the blocklength and error probability constraints.}. 

\begin{thm} \label{thm:noma_case1}
	Suppose that \eqref{monotonicity_condition} is met and that problem \eqref{P-noma-case1} is feasible. The optimal solution to problem \eqref{P-noma-case1} is given by
	\begin{align} \label{optimal_solution_case1}
	\left\{
	\begin{array}{ll}
	m_k^* = D_k,  ~~ {\rm for}~ k=1,2,\\
	\gamma_k^* ~= \Gamma_k(D_k),  ~~ {\rm for}~ k=1,2,\\
	p_1^* ~=  \frac{\gamma_1^*\gamma_2^*}{h_2} + \frac{\gamma_1^*}{h_1}=\frac{\Gamma_1(D_1)\Gamma_2(D_2)}{h_2} + \frac{\Gamma_1(D_1)}{h_1},\\
	p_2^* ~= \frac{\gamma_2^*}{h_2}=\frac{\Gamma_2(D_2)}{h_2},
	\end{array}
	\right.
	\end{align}
where the implicit function $\Gamma_k(\cdot)$ satisfies \eqref{eq_Gamma_P_MOMA}.
\end{thm}

\begin{IEEEproof}
The proof is relegated to Appendix \ref{app:thm1}.
\end{IEEEproof}	

Note that even though the optimal SINR
$\gamma_k^* ~= \Gamma_k(D_k)$ involves implicit function $\Gamma_k(.)$, the inverse $\Gamma^{-1}_k(\gamma_k)$ can be expressed in closed-form as \eqref{gamma_inv} which is due to the fact that \eqref{p1.1} can be viewed as a quadratic equation of $\sqrt{m_k}$ for given $\gamma_k$.
Then it results in the low-complexity Algorithm \ref{alg:bisection} for solving $\gamma_k^*$.

\begin{algorithm}[!tb] \small 
	\caption{Algorithm to find optimal SINR for problem \eqref{P-noma-case1}}\label{alg:bisection}
	\begin{algorithmic}[1]
		\STATE {{\bf Given} the initial values $\Gamma_{\ell k} = 0$, $\Gamma_{uk} = P_{\max}h_k + \delta$ with $\delta > 0$, and the tolerance $\epsilon_0$.}\\
		\WHILE {$\Gamma_{uk} - \Gamma_{\ell k} > \epsilon_0$}
		\STATE {$\bar{\gamma}_k = \frac{1}{2}(\Gamma_{uk} + \Gamma_{\ell k})$.}
		\STATE {Compute $\bar{m}_k=\Gamma_k^{-1}(\bar{\gamma}_k)$ as }
		\!\!\!\!\!\! \begin{align}
		&\!\!\!\!\!\!\!\Bigg[\frac{1}{2\log_2(1\!+\!\bar{\gamma}_k)}\Bigg(\frac{ Q^{-1}(\epsilon_k)}{\ln 2}\sqrt{1\!-\!\frac{1}{(\bar{\gamma}_k\!+\!1)^2}} \nonumber\\
		&\!\!\!+\!\sqrt{\!\!\left(\!1\!-\!\frac{1}{(\bar{\gamma}_k\!+\!1)^2}\!\right)\!\!\left(\!\frac{Q^{-1}(\epsilon_k)}{\ln 2}\!\right)^2 \!\!\!\!+\! 4N_k \log_2 (1\!+\!\bar{\gamma}_k)}\!\Bigg) \!\Bigg]^2 \label{gamma_inv} 
		\end{align}
		\IF {$\bar{m}_k < D_k$ }
		\STATE {\quad Update $\Gamma_{uk} = \bar{\gamma}_k$.}\\
		\ELSE 
		\STATE {\quad Update $\Gamma_{\ell k} = \bar{\gamma}_k$.}\\
		\ENDIF
		\ENDWHILE
		\STATE {{\bf{Output :}}$ \gamma^*_k = \Gamma_k(D_k^*)$}
	\end{algorithmic}
\end{algorithm}

\begin{rem} \label{rem:feasibility}
Note that for given block error rate and latency requirements, problem \eqref{P-noma-case1} may not be feasible due to the limited $P_{\max}$ and the deep channel fadings.
	However, thanks to the obtained closed-form solution of problem \eqref{P-noma-case1} in \eqref{optimal_solution_case1}, its feasibility can be easily checked.
	In particular, from the proof of Theorem \ref{thm:noma_case1}, if
	\begin{equation} \label{eq_feasible_smallh1}
	\frac{\Gamma_1(D_1)\Gamma_2(D_2)}{h_2} + \frac{\Gamma_1(D_1)}{h_1}+\frac{\Gamma_2(D_2)}{h_2}\le P_{\max},
	\end{equation}
	then problem \eqref{P-noma-case1} is feasible under $P_{\max}$ and for channel realizations $(h_1,h_2)$. Otherwise, problem \eqref{P-noma-case1} is infeasible.
\end{rem}

\begin{rem}\label{rem:reliability}
	It is important to point out that
		the overall communication reliability of receiver $k$ is the product of the receiver decoding probability and the feasibility of problem \eqref{P-noma-case1}.
		For instance, assume the probability that \eqref{eq_feasible_smallh1} is not satisfied is $\epsilon_{\rm ifp}$, the overall reliability for receiver 2 should be $(1-\bar{\epsilon}_2)(1-\epsilon_{\rm ifp}) \approx 1 - \bar{\epsilon}_2 -\epsilon_{\rm ifp}$, where a upperbound of $\bar{\epsilon}_2$ is given in \eqref{eq_user2errSmallh1}.
		Note that for given block error rate and latency requirements, the feasibility of the optimization problems is determined by $P_{\max}$ and the random channel realizations. Thus, for a given distribution of the channel gain, the block error rate and $P_{\max}$ should be jointly selected to guarantee the communication reliability of the receivers, which will be studied in the simulation results section.
\end{rem}

\begin{rem}
	{Our results also help to solve the NOMA latency minimization problems, which were considered in \cite{ding-mec-1,ding-mec-2} using Shannon capacities, with FBCs to cope with stringent latency constraints.}  Let's consider the latency minimization problem with $h_1 \le h_2$ as follows
	\begin{subequations} \label{p:latency_min_1}
		\begin{align}
		\min_{m_1,m_2,p_1,p_2} ~~ & m_2\\
		\st ~~~ & F_k(m_k,p_k) = 0, \\
		& \hat{m} \le m_1 \le D_1,~m_1 \le m_2, \label{eq:latecy_m2}\\
		& p_1 + p_2 \le P_{\max}, p_1 \ge 0, p_2 \ge 0,  \label{eq:latecy_power} 
		\end{align}
	\end{subequations}
{where as \eqref{transformation_1}, $F_k(m_k,p_1,p_2)$ is based on \eqref{eq:rate_receiver1} and \eqref{5} for $k = 1,2$ respectively.
		By denoting $m_2^*$ as the optimal latency of user 2 and according to constraint \eqref{eq:latecy_m2}, problem \eqref{p:latency_min_1} can be divided into two cases, i.e., $m_2^* \le D_1$ and $m_2^* > D_1$.
		According to \eqref{eq:rate_receiver1} and \eqref{5}, define the power function as $p_2 = P_2(m_2)$  and $p_1 = P_1(m_1,m_2)$. We have $P_2(m_2)$ is a decreasing function of $m_2$ according to \cite[Proposition 1]{Xu-2016}. Similarly, fix $m_2$, $P_1(m_1,m_2)$ is also a decreasing function of $m_1$. Thus for the case $m_2^* \le D_1$, we have $m_1^* = m_2^*$ from \eqref{eq:latecy_m2}. Note that the optimal $p_1^*$ and $p_2^*$ satisfies that $p_1^* + p_2^* = P_{\max}$, otherwise, $p_1^*$ and $p_2^*$ can be increased accordingly to decrease $m_2^*$ as $P_2(m_2)$ is a decreasing function. Now the optimal $m_2^*$ can be found as follows. Given $m_2$, the corresponding $p_2$ can obtained as Algorithm \ref{alg:bisection}. With $m_1 = m_2$, $p_1$ can be found accordingly. If the power constraint \eqref{eq:latecy_power} is satisfied, $m_2$ can be decreased ; otherwise, $m_2$ should be increased. As a result, the optimal $m_2^*$ can be attained in a bisection manner. For the case that $m_2^* > D_1$, similar approach can be used to find the optimal $m_2^*$.}
\end{rem}

\subsubsection{ Optimal Solutions for Problem \eqref{P-noma-case21} and \eqref{P-noma-case22}} \label{subSec_solution2}
Similar to problem \eqref{P-noma-case1}, the optimal solutions of problem \eqref{P-noma-case21} and \eqref{P-noma-case22} can be obtained by using the monotonicity in Lemma \ref{lem:energy_monotonicity}, which are summarized in the following corollaries.
\begin{cor} \label{cor_noma_case21}
	If condition \eqref{monotonicity_condition} is met, the optimal solution of problem \eqref{P-noma-case21} is given by
	\begin{align}\label{optimal_solution_case2}
	\left\{
	\begin{array}{ll}
	m_k^* = D_k,  ~~ {\rm for}~ k=1,2,\\
	\gamma_k^* ~= \Gamma_k(D_k),  ~~ {\rm for}~ k=1,2,\\
	p_1^* ~= \frac{\gamma_1^* h_2 + \gamma_1^* \gamma_2^* h_1 }{h_1h_2(1 - \gamma_1^* \gamma_2^*)},\\
	p_2^* ~= \frac{\gamma_2^* h_1 + \gamma_1^* \gamma_2^* h_2 }{h_1h_2(1 - \gamma_1^* \gamma_2^*)},
	\end{array}
	\right.
	\end{align}
	whenever it is feasible, where the optimal SINR $\gamma_k^* (k=1,2)$ can be obtained through Algorithm \ref{alg:bisection}.
\end{cor}

\begin{cor} \label{cor_noma_case22}
	If condition \eqref{monotonicity_condition} is met, the optimal solution of problem \eqref{P-noma-case22} is given by
	\begin{align} \label{optimal_solution_case21}
	\left\{
	\begin{array}{ll}
	m_k^* = D_1,  ~~ {\rm for}~ k=1,2,\\
	\gamma_k^* ~= \Gamma_k(D_1),  ~~ {\rm for}~ k=1,2,\\
	p_1^* ~= \frac{\gamma_1^*}{h_1},\\
	p_2^* ~= \frac{\gamma_1^* \gamma_2^*}{h_1} + \frac{\gamma_2^*}{h_2},
	\end{array}
	\right.
	\end{align}
	whenever it is feasible, where the optimal SINR $\gamma_k^*(k=1,2)$ can be obtained through Algorithm \ref{alg:bisection}.
\end{cor}

\begin{IEEEproof}
	The proofs of Corollary \ref{cor_noma_case21} and \ref{cor_noma_case22} are similar to that of Theorem \ref{thm:noma_case1}. Thus we omit them here.
\end{IEEEproof}

It is important to emphasize that, due to the heterogeneous latency requirements ($D_1 < D_2$) of the receivers, the proposed NOMA scheme is conservative and \textit{cannot} achieve the same performance of traditional NOMA schemes that assume perfect SIC \cite{XU-TSP-2017}. Specifically, when $h_1 < h_2$, we have assumed that receiver 2 performs SIC given that the interfering signal from receiver 1 has the same blocklength as the signal of receiver 2.
 However, in fact, there is no interference during the last $D_2 - D_1$ symbols; on the other hand, when receiver 1 performs SIC for $h_1 \ge h_2$, it needs $m_2 = m_1 = D_1 < D_2$. Thus the
blocklength of receiver 2 is limited,
which would incur more energy consumption according to Lemma \ref{lem:energy_monotonicity}. With this consideration, we investigate a novel hybrid transmission scheme in the next section.


\section{Proposed Hybrid NOMA Transmission Schemes} \label{sec_hybrid}

In this section, we introduce data splitting with time domain power allocation for NOMA and propose a hybrid scheme that incorporates the NOMA in the previous section and TDMA as two special cases. In particular, the data packet for user 2 is split into two parts, where the first part has $N_{21}$ bits and the second has $N_{22}$ bits and $N_{21}+N_{22}=N_2$. The $N_{21}$ bits are encoded into $m_{21}$ symbols, and combined with the encoded symbols for user 1 using the non-orthogonal super-position coding; whereas the rest $N_{22}$ bits are encoded into $m_{22}$ symbols and scheduled in the non-overlapping time slots. Note that when $N_{21} = 0$, the hybrid scheme degrades into the TDMA studied in \cite{Xu-2016}; while when $N_{22} = 0$, the hybrid scheme degrades into the NOMA in Section \ref{sec_NOMA}. As in Section \ref{subsec_largerh2}, according to the different channel conditions at receivers, we study the problem by considering the following cases.

\begin{figure}[!tp]
	\centering
	\includegraphics[scale=0.6]{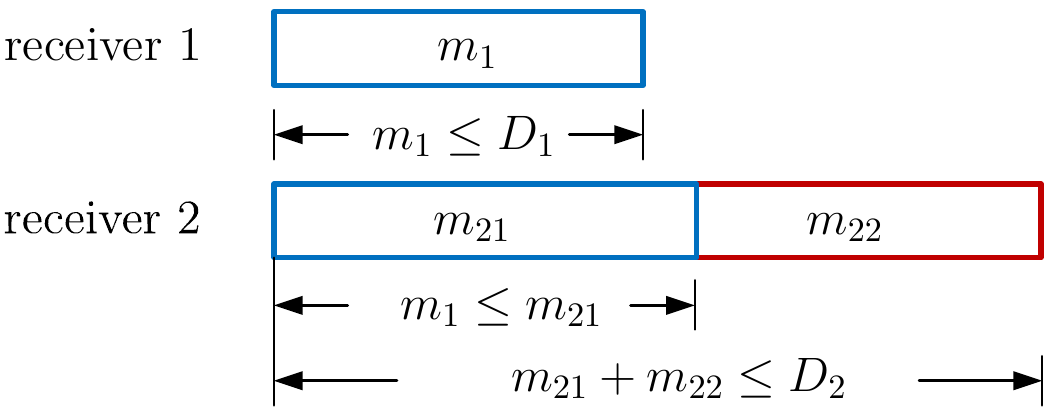}\\
	\caption{Hybrid transmission scheme where the packet of receiver 2 is split into two parts which are scheduled with $m_{21}$ and $m_{22}$ symbols respectively. Here $h_1 \leq h_2$ and receiver 2 performs SIC.} \label{hybrid_scheme_case1}
\end{figure}

\noindent
\textbf{Case I} ($h_1 \leq h_2$ with SIC at receiver 2) : In this case, the transmission scheme is sketched in Fig. \ref{hybrid_scheme_case1} and receiver 2 performs SIC. The transmitter first transmits the $N_1$ bits of receiver 1 and $N_{21}$ bits of receiver 2 by using the NOMA scheme. The transmit signal is $\sqrt{p_1} x_1 + \sqrt{p_{21}} x_{21}$, where $x_{21}$ and $p_{21}$
are the unit-power coded symbols and corresponding allocated power for user 2, respectively. Also we have $p_1 + p_{21} \le P_{\max}$.

\begin{figure}[!tp]
	\centering
	\includegraphics[scale=0.6]{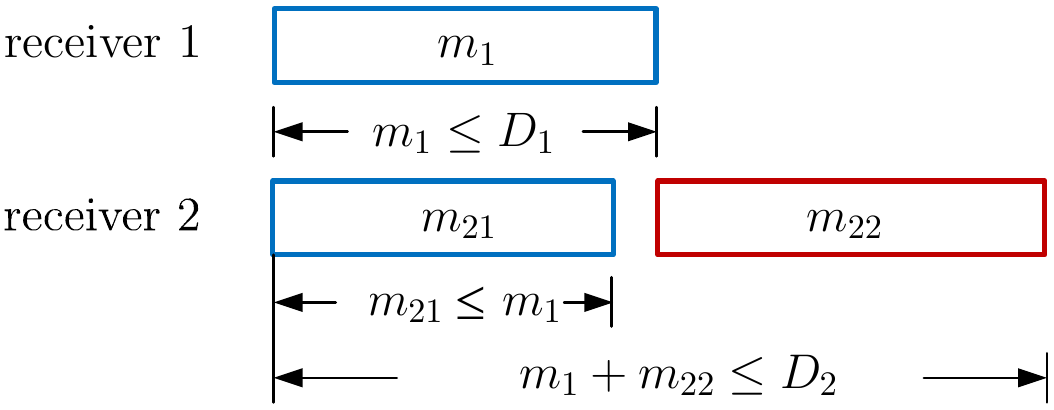}\\
	\caption{Hybrid transmission scheme when $h_2 \le h_1$ and receiver 1 performs SIC.} \label{fig:hybrid_scheme_case2}
\end{figure}

For user 1, the $N_1$ bits are encoded with a FBC of length $m_1$ and the achievable rate is same as \eqref{eq:rate_receiver1} with SINR
$\gamma_1 = \frac{p _{1} h_1}{p_{21}h_1 + 1}$.
For receiver 2, it can cancel the interference from user 1 using the received signal from the first $m_{21}$ received symbols since $\frac{p_{1}h_2}{p_{21}h_2 + 1} > \gamma_1.$
After that, receiver 2 decodes its own information with SINR $\gamma_{21} = p_{21}h_2$,
and from the corresponding achievable rate $\frac{N_{21}}{m_{21}}$ satisfies
\begin{align}
 \frac{N_{21}}{m_{21}} \!=\! \log_2(1 \!+\! \gamma_{21}) \!-\!\!
\sqrt{\!\!\frac{1}{m_{21}}\!\!\left(\!1 \!-\! \frac{1}{(\gamma_{21} \!+\!1)^2}\!\right)}
\frac{Q^{-1}\!(\epsilon_{21})}{\ln2}.
\end{align}
Remind that receiver 2 needs to receive all information symbols of receiver 1 to perform SIC, thus we require that
\begin{align}
m_{21}\ge m_1.
\end{align}
Once the transmission of the first $m_{21}$ symbols is complete, the transmitter starts to deliver the rest $N_{22}$ bits solely for user 2 using $m_{22}$ symbols and power $p_{22}$. The achievable rate $\frac{N_{22}}{m_{22}}$ satisfies
\begin{align}
	 \frac{N_{22}}{m_{22}} \!=\! \log_2(1 \!+\! \gamma_{22}) \!-\!\!
	\sqrt{\!\!\frac{1}{m_{22}}\!\!\left(\!1 \!-\! \frac{1}{(\gamma_{22} \!+\!1)^2}\!\right)}
	\frac{Q^{-1}\!(\epsilon_{22})}{\ln2}.
\end{align}
where $\gamma_{22} = p_{22} h_2$. Notice that the SIC at receiver 2 is successful only when the interference (from user 1's message) is perfectly subtracted as well as the $N_{21}$ and $N_{22}$ bits are both successfully decoded. Also for the decoding of user 1's codeword, the successful probability at stronger receiver 2 will be larger than the one at weaker receiver 1, i.e., $1-\epsilon_1$. 
Then the overall block error rate for receiver 2, i.e., $\epsilon_{2}$, is upper-bounded by
		\begin{equation} \label{eq_user2errHybrid}
		\epsilon_2
		\leq 1 - (1-\epsilon_1) (1-\epsilon_{21})(1 - \epsilon_{22}).		
		\end{equation}
With slightly abuse of notations, the implicit SINR function in \eqref{eq_Gamma_P_MOMA} is denoted as
$
\gamma_k \triangleq \Gamma_k(N_k,m_k).
$
Therefore, the energy minimization problem can be formulated as follows
\begin{subequations} \label{P-hybrid-shceme-case1}
	\begin{align}
	\min_{\substack{N_{21},N_{22},\\m_1,m_{21},m_{22},\\p_1,p_{21},p_{22}}} & m_1 p_1 + m_{21}p_{21} + m_{22} p_{22} \\
	\st ~~~~
	& \!\!\!\!\!\!\gamma_1 =  \frac{p_1 h_1}{p_{21}h_1 + 1} =  \Gamma_1(N_1,m_1),\label{FBC-1} \\
	& \!\!\!\!\!\!\gamma_{21} = p_{21} h_2 = \Gamma_{21}(N_{21},m_{21}),\label{FBC-2}\\
	& \!\!\!\!\!\!\gamma_{22} = p_{22} h_2 = \Gamma_{22}(N_{22},m_{22}),\label{FBC-3} \\
	& \!\!\!\!\!\!m_1 \!\le\! \min\{D_1, m_{21}\}, m_{21} \!+\! {\mathds{1}}(N_{22})m_{22} \!\le\! D_2,\nonumber \\
	& \!\!\!\!\!\!m_1,m_{21},m_{22} \ge \hat{m},\label{LatencyConstraints}\\
	& \!\!\!\!\!\!p_1 \!+\! p_{21} \!\le\! P_{\max}, p_{22} \!\le\! P_{\max}, p_1,p_{21},p_{22} \!\ge\! 0, \label{PowerConstraints} \\
	& \!\!\!\!\!\!N_{21} + N_{22} = N_2, ~0 \le N_{21} \le N_2.\label{packet-size2}
	\end{align}
\end{subequations}
where $\mathds{1}(N_{22} \neq 0) = 1$ denotes the indicator function. 
Note that when $N_{21}=0$ problem \eqref{P-hybrid-shceme-case1} degrades to energy minimization with TDMA; and when $N_{22} = 0$, problem \eqref{P-hybrid-shceme-case1} degrades into \eqref{P-noma-case1} with pure NOMA. Since \eqref{P-noma-case1} has been already solved in Section \ref{Sec_solution}, we only need to focus on the cases of $0 \leq N_{21} \le N_2 -1$ (thus $N_{22} \geq 1$) in problem \eqref{P-hybrid-shceme-case1}.\\
\noindent
\textbf{Case 2} ($h_2 < h_1$ with SIC at receiver 1): In this case, receiver 1 performs SIC to cancel the interference from user 2 and the transmission scheme is depicted in Fig. \ref{fig:hybrid_scheme_case2}.
To satisfy the requirement of SIC at receiver 1, the latency constraints are given by
	\begin{align}\label{delay_constraints}
	& m_{21} \le m_1 \!\le\! D_1, m_{1} + m_{22} \!\le\! D_2, m_1,m_{21}, m_{22} \ge \hat{m}.
	\end{align}
The energy minimization problem can be formulated as
\begin{subequations}\label{P-hybrid-shceme-case2}
	\begin{align}
	\min_{\substack{N_{21},N_{22},\\m_1,m_{21},m_{22},\\p_1,p_{21},p_{22}}} ~& m_1 p_1 + p_{21}m_{21} + p_{22}m_{22} \\
	\st ~~~~~& \eqref{FBC-3}, \eqref{PowerConstraints}, \eqref{packet-size2}, \eqref{delay_constraints},\\
	& \gamma_1 = p_1 h_1 = \Gamma_1(N_1,m_1),\\
	& \gamma_{21} = \frac{p_{21} h_2}{p_{1}h_2 + 1} = \Gamma_{21}(N_{21},m_{21})
	\end{align}
\end{subequations}

Here we should point out that problems \eqref{P-hybrid-shceme-case1} and \eqref{P-hybrid-shceme-case2} are more challenging to solve compared to problems \eqref{P-noma-case1}, \eqref{P-noma-case21} and \eqref{P-noma-case22} in Section \ref{sec_NOMA} due to the following reasons. First, the blocklengths $m_{21}$ and $m_{22}$ of user 2 in the two transmission stages are coupled in the constraints, consequently, the monotonicity of the energy function {\textit{cannot}} be used to attain the solution directly as in Section \ref{sec_NOMA}. Second, now the integer packet sizes $N_{21}$ and $N_{22}$ of user 2 are the additional optimization variables, which also complicates the constraints  \eqref{FBC-2} and \eqref{FBC-3}. To solve these problems efficiently, a concave approximation of the FBC capacity formula is provided next.

\subsection{Convex Approximation of the normal approximation of FBC capacity formula}\label{sec_covexapprox}
We start from the convexity analysis of the normal approximation of FBC capacity formula as in the upcoming Proposition \ref{prop:convexity_FBC} and then propose a concave approximation of the normal approximation of FBC capacity formula in \eqref{eq:modified_fbc}, with given blocklength and error probability. The inverse of proposed approximation function (33) is also convex. For the simplicity of notation, we remove the subindex of all variables and let $x$ denote the \SINR. The normal approximation of FBC capacity formula then becomes
\begin{align}\label{func_F1}
\frac{N}{m} = \ln(1+x) - \sqrt{\frac{1}{m}\left(1-\frac{1}{(1+x)^2}\right)}\frac{Q^{-1}(\epsilon)}{\ln(2)}.
\end{align}
With given blocklength $m$ and error probability $\epsilon$, by letting $ a = \frac{Q^{-1}(\epsilon)}{\sqrt{m}\ln 2 }$, we define $f(x)$ as
\begin{align} \label{func:fx}
f(x) \triangleq \ln(x+1) - a \frac{\sqrt{x(x+2)}}{x+1}
\end{align}
Also, given $m$, we define SINR function with respect to packet size as $\Gamma(N,m) = f^{-1}(\frac{N}{m})$.
Then we have the following proposition, with the proof relegated to Appendix \ref{app2}, as
\begin{prop} \label{prop:convexity_FBC}
Let constant $\beta \triangleq g(x_0)= g_2(x_0)$, where $x_0 = 0.6904$ is the positive solution of equation $g_2(x) = g(x)$ with
\begin{subequations}
	\begin{align}
	g_2(x) &\triangleq \frac{(x+1)(x(x+2))^{\frac{3}{2}}}{3x^2+6x+1} \\
	g(x) &\triangleq \frac{(x+1)\ln(x+1)}{\sqrt{x(x+2)}}.
	\end{align}
\end{subequations}
For a given $a$, the convexity of $f(x)$ in \eqref{func:fx} is given by
	\begin{align} \label{eq:convexity_of_f(x)}
	\begin{cases}
	{\rm if}~a \!>\! \beta, f(x) {\rm~is~concave~and~increasing} ~{\rm for}~x\!>\!g^{-1}(a),\\
	{\rm if}~a \!\le\! \beta, \\
	\begin{cases}
	f(x) {\rm ~is~concave~and~increasing}  ~{\rm for} ~ x\!>\!g_2^{-1}(a),\\
	f(x) {\rm ~is~convex~and~increasing}  ~{\rm for} ~ g^{-1}(a) \!\le\! x \!\le\! g_2^{-1}(a).\\
	\end{cases}
	\end{cases}
	\end{align}
\end{prop}
\begin{IEEEproof}
	The proof is relegated to Appendix \ref{app2}
	\end{IEEEproof}

\noindent Note that only positive $f(x)$ is meaningful by definition, so $x$ must be larger than $g^{-1}(a)$ from \eqref{func:fx}. Besides, whenever $f(x)$ is concave in $x$, $\Gamma(N,m)$ is convex in $N$.

\begin{figure}[!tp]
	\centering
	\includegraphics[scale=0.64]{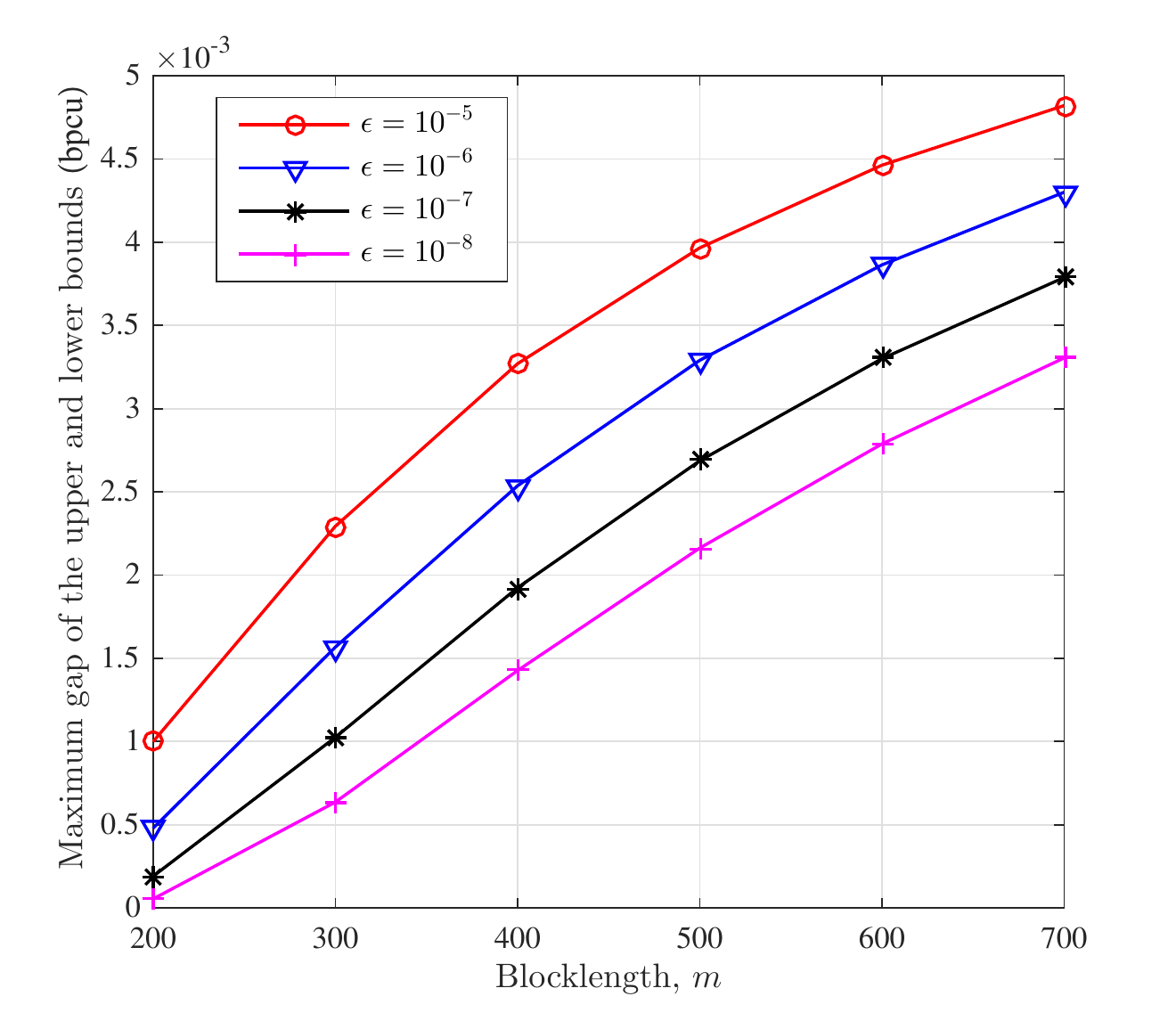}\\
	\caption{The maximum gap between upper bound and lower bound with different blocklength $m$ and error probability $\epsilon$, where bpcu is the abbreviation of bits per channel use.} \label{fig:capacity_formulas}
\end{figure}

\begin{figure}[!tp] \centering
	\subfigure[FBC capacity formula function $f(x)$ for the case $a \le \beta$ in Proposition \ref{prop:convexity_FBC}.]{
		\includegraphics[width=0.4\textwidth]{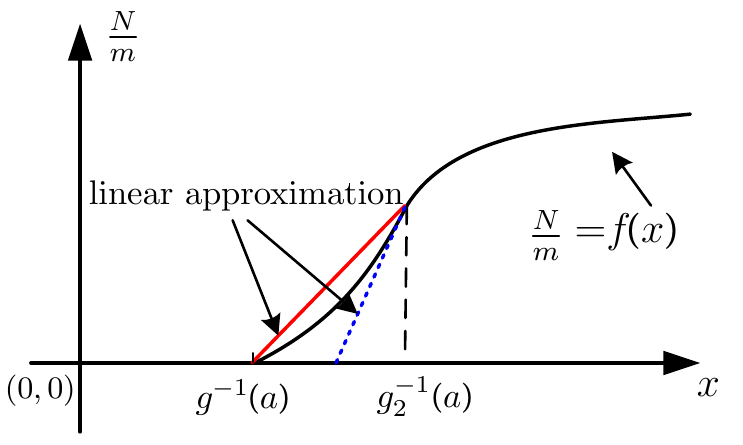}\label{fig:rate_depiction}}
	\hspace{15mm}
	\subfigure[The inverse function of $f(x)$, or the SINR function $\Gamma(N,m)$ with fixed $m$, for the cases of $a \le \beta$.]{
		\includegraphics[width=0.4\textwidth]{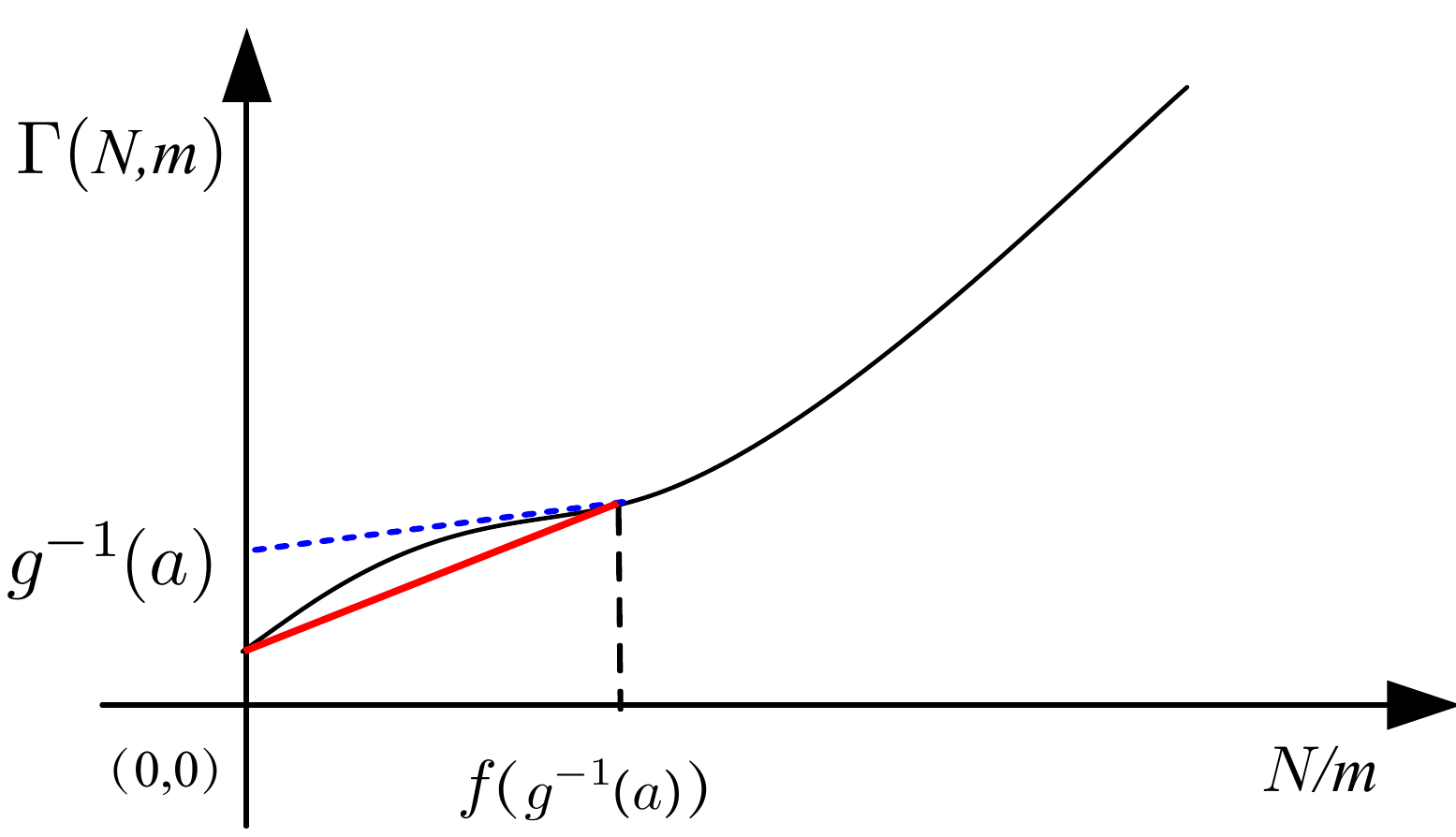} \label{fig:gamma_depiction}}\\
	\caption{(Schematic) Illustration of the convexity of the normal approximation of FBC capacity formula function $f(x)$ and its inverse function.}
	\label{fig:g_g2} 
\end{figure}

Proposition \ref{prop:convexity_FBC} shows that $f(x)$ ($\Gamma(N,m)$) is in fact not always concave (convex). To overcome this problem, as shown in Fig. \ref{fig:rate_depiction},
we consider approximating $f(x)$ in the interval $g^{-1}(a) \le x < g_2^{-1}(a)$ by taking a linear upper bound and lower bound. Specifically, since $f(x)$ is convex in the interval $g^{-1}(a) \le x < g_2^{-1}(a)$ for $a \le \beta$, the linear function
\begin{align}\label{eq:linearization_ub}
f_u(x) = \frac{f\left(g_2^{-1}(a) \right)}{g_2^{-1}(a) - g^{-1}(a)} \left(x - g^{-1}(a)\right);
\end{align}
is an upper bound of $f(x)$; the first-order Tylor expansion of $f(x)$ at $x = g_2^{-1}(a)$, i.e.,
\begin{align}\label{eq:linearization_lb}
f_{\ell}(x) = f^{\prime}\left(g_2^{-1}(a)\right)\left(x-g_2^{-1}(a)\right) + f\left(g_2^{-1}(a)\right),
\end{align}
is a lower bound of $f(x)$.
Notice that the maximum gap happens when the lower bound equals to zero. To examine the tightness of the upper and lower bound, in Fig. \ref{fig:capacity_formulas},  we plot the gap under different blocklength $m$ and error probability $\epsilon$. One can see that the approximations are tight.

Although both of the linear approximations are quite tight from Fig. \ref{fig:capacity_formulas}, we will use the lower-bound approximation \eqref{eq:linearization_lb}. First, the lower-bound approximation can be treated as an achievable rate. Second, by using the lower-bound approximation, the approximated FBC capacity formula $f(x)$ is a concave function. Then the corresponding SINR function $\Gamma_k(\cdot)$ in  
Problems \eqref{P-hybrid-shceme-case1} and \eqref{P-hybrid-shceme-case2},
is increasing and convex in the (normalized) number of data bits $N/m$ \cite[Proposition 2]{mila-2008}. An example is shown in Fig \ref{fig:gamma_depiction}. In the meanwhile, with the upper-bound approximation, the approximated FBC capacity formula is only quasi-concave, and thus the convexity of SINR functions cannot be guaranteed. Finally, by using \eqref{eq:linearization_lb}, the approximation of FBC capacity formula with given blocklength $m$ and error probability \eqref{eq:convexity_of_f(x)} reads
\begin{align} \label{eq:modified_fbc}
&\frac{N}{m} =f_M(x) = \nonumber\\
&\begin{cases}
\ln(x+1) \!-\! a \frac{\sqrt{x(x+2)}}{x+1} , \!\!\!\!& {\rm if}
\begin{cases}
a \!>\! \beta~\&~x \ge g^{-1}(a)\\
a \le \beta~\&~x \ge g_2^{-1}(a)
\end{cases}\\
\eqref{eq:linearization_lb}, & \!\!\!\!\!\!\!\!\!\!\!\!\!\!\!\!\!\! {\rm if}~a \!\le\! \beta ~\&~g^{-1}(a) \le x \!<\! g_2^{-1}(a).
\end{cases}
\end{align}
where $x$ is the SINR while $a$ and $\beta$ are given in Proposition  \ref{prop:convexity_FBC}.

\setcounter{TempEqCnt}{\value{equation}} 
\setcounter{equation}{38} 
\begin{figure*}[!t] 
	\begin{subequations}\label{eq:39}
		\begin{align}
		& \frac{D_1 \Gamma_1(\!N_1,D_1\!)}{h_1} \!+\! \frac{D_1 \Gamma_1(N_1,D_1\!)\Gamma_{21}(N_{21},m_{21}\!)}{h_2}\!+\!\frac{m_{21} \Gamma_{21}(N_{21},m_{21}\!)}{h_2} + \frac{(D_2 - m_{21})\Gamma_{22}(N_2 - N_{21},D_2 - m_{21})}{h_2} \\
		=& \frac{(m_{21}\!\!+\! D_1\Gamma_1(N_1,D_1)) \Gamma_{21}(N_{21},m_{21})}{h_2} \!+\! \frac{(D_2 \!-\! m_{21})\Gamma_{22}(N_2 \!-\! N_{21},D_2 \!-\! m_{21})}{h_2} \!+\!  \frac{D_1 \Gamma_1(N_1,D_1)}{h_1}.
		\end{align}
	\end{subequations}
	\hrulefill	
\end{figure*}

\subsection{Solving Problems \eqref{P-hybrid-shceme-case1} and \eqref{P-hybrid-shceme-case2} with Approximation \eqref{eq:modified_fbc}}
Now we turn to solve the mixed-integer problems \eqref{P-hybrid-shceme-case1} and \eqref{P-hybrid-shceme-case2}. Since \eqref{P-hybrid-shceme-case1} and \eqref{P-hybrid-shceme-case2} have similar structures, we  focus on problem \eqref{P-hybrid-shceme-case1}. We first provide a solver based on exhaustive linear search as a benchmark, and then present our more efficient one based on Lemma \ref{lem:energy_monotonicity} and approximation \eqref{eq:modified_fbc}. \\
\textbf{Benchmark : Solver using exhaustive linear search} :
 Note that the SINR constraints and transmit power constraints in \eqref{PowerConstraints} actually restrict the packet size and blocklengths to a subset that
 \setcounter{TempEqCnt}{\value{equation}} 
 \setcounter{equation}{34}
\begin{align}
\mathcal{H} \triangleq \Big\{&N_{21},m_1,m_{21},m_{22} ~\Big| \nonumber\\
&p_1 = \frac{\Gamma_1(N_1, m_1) \Gamma_{21}(N_{21},m_{21})}{h_2} \!+\! \frac{\Gamma_1(N_1,m_1)}{h_1}, \nonumber\\
&p_{21} \!=\!  \frac{\Gamma_{21}(N_{21},m_{21})}{h_2}, p_{22} \!=\!  \frac{\Gamma_{22}(N_2 \!-\! N_{21},m_{22})}{h_2},\nonumber\\
&p_1 + p_{21} \le P_{\max}, ~ p_{22} \le P_{\max}, \nonumber \\
&{\rm and}~ p_1, p_{21}, p_{22} \ge 0{\rm ~are ~satisfied} \Big\}. \label{eq_3Dset}
\end{align}
Thus problem \eqref{P-hybrid-shceme-case1} can be rewritten as
\begin{subequations} \label{P-relaxation1}
	\begin{align}
	\min_{\substack{N_{21},m_1,\\m_{21},m_{22}}}&\frac{m_1 \Gamma_1\!(\!N_1,m_1\!)}{h_1} + \frac{m_{1}\Gamma_{1}\!(\!N_1,m_1\!)\Gamma_{21}\!(N_{21},m_{21}\!)}{h_2}  +\nonumber\\ &\frac{m_{21}\Gamma_{21\!}(\!N_{21},m_{21}\!)}{h_2} \!+\! \frac{m_{22} \Gamma_{22}(\!N_2 \!-\! N_{21},m_{22}\!)}{h_2}\label{energy}\\
	\st~~
	&m_1,m_{21},m_{22} \ge \hat{m},~ m_{21} + m_{22} = D_2, \label{eq:blocklength_hybrid_r2}\\
	&m_1 \le \min \{m_{21}, D_1\}, \label{latency_hybrid_11}\\
	& N_{21} + N_{22} = N_2, ~0 \leq N_{21} \le N_2-1, \\
	& (N_{21},m_1,m_{21},m_{22}) \in \mathcal{H}, \label{eq:m_subset}
	\end{align}
\end{subequations}
where the ``$=$'' in \eqref{eq:blocklength_hybrid_r2} is obtained from Lemma \ref{lem:energy_monotonicity}. 
It can be readily verified that, by ignoring constraint \eqref{eq:m_subset}, problem \eqref{P-relaxation1} is fully determined by variables $N_{21}$, $m_{1}$ and $m_{21}$. Thus the remaining problem can be solved to its global optimal solutions by using a three-dimensional (3-D) exhaustive linear search method.
For each searching point, one only need to check whether it is in the set of constraint \eqref{eq_3Dset} to avoid infeasible power allocation, as in \eqref{eq:m_subset}.

The complexity of the 3-D exhaustive search Algorithm is determined by the
linear searching of $N_{21}$ with complexity order $\mathcal{O} (N_2)$, those of $m_1$ and $m_{21}$ with complexity order $\mathcal{O}\Big(\frac{1}{2}\big(2D_2-3\hat{m}+2 - \min\{D_1,D_2 - \hat{m}\}\big)\big(\min\{D_1, D_2-\hat{m}\} + 2 -\hat{m}\big) \Big) $.
Besides, in each iteration, it contains $3$ times bisection search to find the optimal SINRs with complexity order of
$\mathcal{O}\left(3\log\left(\frac{\max_k\{h_k P_{\max}\}}{\epsilon_0}\right)\right)$ where $\epsilon_0 > 0$ is the desired accuracy.
Summarily, the total complexity order is given by
$\mathcal{O}\Big(\frac{3N_2}{2}\big(2D_2-3\hat{m}+2 - \min\{D_1,D_2 - \hat{m}\}\big) \big(\min\{D_1, D_2-\hat{m}\} + 2 -\hat{m}\big) \log\left(\frac{\max_k\{h_k P_{\max}\}}{\epsilon_0}\right) \Big)$.
Considering the high computational complexity of the 3-D linear search, we seek a more efficient way to solve problem \eqref{P-relaxation1}.

\noindent \textbf{Solver based on Convex Approximation} : Now we present the solver which is more efficient  than the previous one using 3-D linear search. 
Firstly, for either $m_{21}$ being shorter than $D_1$ or not, problem
\eqref{P-relaxation1} can be decoupled into two subproblems. Though it seems that each of them still need two-dimensional linear search, with the aid from the convex approximation of the normal approximation of FBC capacity formula, the search range can be significantly reduced by golden section search \cite{Book_Golden_Section_Search}. In fact, we only need one-dimension exhaustive search (on $m_{21}$) for the solver summarized in Algorithm \ref{alg:exhaustive_search_gss}. More specifically, based on \eqref{latency_hybrid_11} and Lemma \ref{lem:energy_monotonicity}, problem \eqref{P-relaxation1} can be decoupled as cases (a) and (b) in the following \\
\noindent a) $m_{21} < D_1$ : In this case, we have $m_1 = m_{21}$ from Lemma \ref{lem:energy_monotonicity} and thus
the objective function of problem \eqref{P-relaxation1} becomes
\begin{align}
&\frac{m_{21}\Gamma_{1}(N_{1},m_{21})}{h_1} + \frac{m_{21}\Gamma_1(N_1,m_{21})\Gamma_{21}(N_{21},m_{21})}{h_2}+\nonumber\\
&\frac{m_{21}\Gamma_{21}(N_{21},m_{21})}{h_2}\!+\! \frac{(D_2 \!-\! m_{21})\Gamma_{22}(N_2 \!-\! N_{21},D_2 \!-\! m_{21})}{h_2},
\end{align}
and by ignoring constraint \eqref{eq:m_subset}, problem \eqref{P-relaxation1} becomes
\begin{subequations}\label{P-relaxation1_case1}
	\begin{align}
	\min_{\substack{N_{21},m_{21}}}~& \frac{m_{21}\Gamma_{1}(N_{1},m_{21})}{h_1} + \nonumber \\
	&\frac{\left(m_{21}\Gamma_1(N_1,m_{21}) + m_{21}\right)\Gamma_{21}(N_{21},m_{21})}{h_2} + \nonumber \\
	&\frac{(D_2 - m_{21})\Gamma_{22}(N_2 - N_{21},D_2 - m_{21})}{h_2}\\
	\st~~ 
	& \hat{m} \le D_2 -m_{21} , ~ \hat{m} \le m_{21} \le D_1, \\
	& 0 \leq N_{21} \le N_2-1.
	\end{align}
\end{subequations}

\noindent b) $m_{21} \ge D_1$ : In this case, we have $m_1 = D_1$ from Lemma \ref{lem:energy_monotonicity} and the objective function of problem \eqref{P-relaxation1} becomes \eqref{eq:39}.
Therefore by ignoring constraint \eqref{eq:m_subset}, problem \eqref{P-relaxation1} is equivalent to
\setcounter{equation}{39}
\begin{subequations} \label{P-relaxation1_case2}
	\begin{align}
	\min_{\substack{N_{21},m_{21}}} ~&\Big(m_{21} + D_1\Gamma_1(N_1,D_1)\Big) \Gamma_{21}(N_{21},m_{21}) + \nonumber\\
	& (D_2 \!- \!m_{21})\Gamma_{22}(N_2 \!-\! N_{21},D_2 \!-\! m_{21}) \\
	\st~~ 
	& \hat{m} \le D_2 - m_{21},~ D_1 \le m_{21}, \\
	& 0 \leq N_{21} \leq N_2-1.
	\end{align}
\end{subequations}
Notice that problem \eqref{P-relaxation1_case1} and \eqref{P-relaxation1_case2} are determined by variables $m_{21}$ and $N_{21}$, indicating that problem \eqref{P-relaxation1} can be solved by 2-D linear search of $m_{21}$ and $N_{21}$. Remind that we decouple problem \eqref{P-relaxation1} based on the value of $m_{21}$, thus we do linear search of $m_{21}$ first and then find the optimal $N_{21}$.
Specifically, for given $m_{21}$, problem \eqref{P-relaxation1_case1} degrades into
\begin{subequations}\label{P-relaxation1_case1_fix_m21}
	\begin{align}
	\min_{\substack{N_{21}}} ~& \left(m_{21}\Gamma_1(N_1,m_{21}) + m_{21}\right)\Gamma_{21}(N_{21},m_{21}) + \nonumber\\
	&(D_2 - m_{21})\Gamma_{22}(N_2 - N_{21},D_2 - m_{21})\label{energy1}\\
	\st~
	&0 \leq N_{21} \leq N_2-1.
	\end{align}
\end{subequations}
and problem \eqref{P-relaxation1_case2} degrades into
\begin{subequations} \label{P-relaxation1_case2_fix_m21}
	\begin{align}
	\min_{\substack{N_{21}}} ~& \Big(m_{21} + D_1\Gamma_1(N_1,D_1)\Big) \Gamma_{21}(N_{21},m_{21}) + \nonumber \\
	& (D_2 - m_{21})\Gamma_{22}(N_2 - N_{21},D_2 - m_{21}) \label{energy2}\\
	\st~
	&0 \leq N_{21} \leq N_2-1.
	\end{align}
\end{subequations}

With convex approximations in section \ref{sec_covexapprox}, Problem \eqref{P-relaxation1_case1_fix_m21} and \eqref{P-relaxation1_case2_fix_m21} can be solved by bisection search approach which is more efficient than linear searching of $N_{21}$. In particular, with the aid of \eqref{eq:modified_fbc} and \cite[Proposition 2]{mila-2008}, we have the convex approximation of SINR $\Gamma_{21}(N_{21},m_{21})$, which we denote as $\hat{\Gamma}_{21}(N_{21},m_{21})$. It can be verified that if $\hat{\Gamma}_{21}(N_{21},m_{21})$ is convex in $N_{21}$, then  $\hat{\Gamma}_{22}(N_2-N_{21},D_2 - m_{21})$ is also convex in $N_{21}$, and then the low complexity golden section search method \cite{Book_Golden_Section_Search}
can be modified to find the optimal integer $N_{21}$ in \eqref{P-relaxation1_case1_fix_m21} and \eqref{P-relaxation1_case2_fix_m21}.
The remaining challenge is that we still cannot have an explicit expression of  $\hat{\Gamma}_{21}(N_{21},m_{21})$ due to the implicit and complex structure of it.
Thanks to the monotonicity with respect to $N_k$, the  bisection search algorithm as in Algorithm \ref{alg:bisection} can be used to find approximated $\hat{\Gamma}_{21}(N_{21},m_{21})$ with given $m_{21}$.
The overall solver with convex approximation is described in Algorithm \ref{alg:exhaustive_search_gss}, which consists of one dimension exhaustive line search of $m_{21}$ and one dimension golden section search of $N_{21}$.


{\begin{algorithm}[!t]\smaller[1]
		\caption{Proposed convex approximation based algorithm for problem \eqref{P-relaxation1}}\label{alg:exhaustive_search_gss}
		\begin{algorithmic}[1]
			\STATE {{\bf Given} system parameters $N_1,~N_2$, $D_1$, $D_2$, $\epsilon_1,~ \epsilon_{21}$, $\epsilon_{22}$ and accuracy $\epsilon_0$.}\\
			\FOR{ $m_{21} = \hat{m}:D_2-\hat{m}$}		
			\STATE{Given $A = 0.618$ and set $N_{\min} = 0,~N_{\max} = N_2-1$;}
			\WHILE{ $N_{\max} - N_{\min} \ge \epsilon_0$}
			\STATE{Set $N_{\ell} = (1-A)(N_{\max} - N_{\min})$ and $N_{u} = A(N_{\max}-N_{\min})$.}
			\STATE{Calculate $\Gamma_{21}(N_{21},m_{21})$ with $N_{21} = N_\ell$ or $N_u$ based on \eqref{eq:linearization_lb} or Algorithm 1.}
			\STATE{Calculate $E_{\ell} = E(N_{\ell},m_{21})$ and $E_{u} = E(N_u,m_{21})$ based on \eqref{energy1} or \eqref{energy2}.}
			\IF{$E_{\ell} \ge E_u$}
			\STATE {Update $N_{\min} = N_{\ell}$;}
			\ELSE
			\STATE{Update $N_{\max} = N_u$.}
			\ENDIF
			\ENDWHILE
			\STATE{Let $\bar{N}_{\ell} = \left\lfloor N_{\ell}\right\rfloor$ and $\bar{N}_{u} = \left\lceil N_{u}\right\rceil$.}
			\STATE{The optimal $N_{21}$ is the one of $\bar{N}_{\ell}$ and $\bar{N}_u$ that minimize the consumed energy.}
	    	\IF {The power constraint in \eqref{eq:m_subset} is satisfied}
			\STATE {Calculate the consumed energy and store $m_{21}$ and $N_{21}$.}
			\ELSE 
			\STATE {Update $m_{21} = m_{21}+1$, and repeat step 3-15.}
			\ENDIF
			\ENDFOR
			\STATE {{\bf Output :} The solutions $m_{21}^*$ and $N_{21}^*$ that minimizes the consumed energy.}
		\end{algorithmic}
\end{algorithm}}


The computational complexity of Algorithm \ref{alg:exhaustive_search_gss} is shown as follows. Given the latency requirement of receiver 2, $D_2$,
the outer search of $m_{21}$ needs $\mathcal{O}(D_2 -2\hat{m})$ rounds to find the optimal $m_{21}$. In each round,
the golden section search will be applied to find the optimal $N_{21}$ with a complexity order of $\mathcal{O}\left(\log_\phi\left(\frac{N_2}{\epsilon_0}\right)\right)$,
where $\phi = 1/A$ and $A$ is the golden section search parameter.
and at most 4 times bisection search to calculate the corresponding $\Gamma_{21}(N_{21},m_{21})$. Thus
the worse case complexity order in each round of golden section search to find $N_{21}$ is given by $\mathcal{O}\left(4\log_2\left(\frac{\max_k\{P_{\max} h_k\}}{\epsilon_0}\right)\right)$. In summary,
the total complexity of algorithm \ref{alg:exhaustive_search_gss} is bounded by $\mathcal{O}\Big(4(D_2 -2\hat{m})\log_\phi\left(\frac{N_2 }{\epsilon_0}\right)\log_2\left(\frac{\max_k\{P_{\max} h_k\}}{\epsilon_0}\right) \Big)$. Compared to the 3-D search based algorithm, the computation complexity of algorithm \ref{alg:exhaustive_search_gss} is reduced dramatically by transforming the 3-D exhaustive search to a one dimension exhaustive search plus one dimension golden section search approach.

Finally, we present a solver for problem \eqref{P-hybrid-shceme-case2}. Similar to that in problem \eqref{P-hybrid-shceme-case1}, we remove the power constraints in problem \eqref{P-hybrid-shceme-case2} and solve it with the linear search method. Based on the monotonicity of the energy function in Lemma \ref{lem:energy_monotonicity}, the optimal $m_{21}$ satisfies $m_{21}^* = m_1$. Thus problem \eqref{P-hybrid-shceme-case2} can be solved by 2-D linear search of $m_1$ and $N_{21}$. Therefore, for any given $m_1$,  problem \eqref{P-hybrid-shceme-case2} is equivalent to
\begin{align}\label{P-hybrid-case2-relaxation}
\min_{0 \leq N_{21} \leq N_2-1} ~& \frac{m_1\Gamma_1(N_1,m_1)\Gamma_{21}(N_{21},m_1) }{h_1}+\nonumber\\
 &\frac{m_1\Gamma_{21}(N_{21},m_1)}{h_2}+\frac{\Gamma_{22}(N_2 - N_{21},D_2-m_1)}{h_2}
\end{align}
Same as problem \eqref{P-relaxation1_case1_fix_m21} and \eqref{P-relaxation1_case2_fix_m21}, by using the convex approximation of the normal approximation of FBC capacity formula \eqref{eq:modified_fbc}, problem \eqref{P-hybrid-case2-relaxation} can be efficiently solved with the bisection search method.


\section{Simulation Results}\label{sec_simu}


In this section, simulation results are given to compare the performance of NOMA and hybrid scheme with that of the TDMA under FBC\footnote{One may consider using FDMA other than TDMA for URLLC. The benefits of FDMA compared with TDMA mainly come from the orthogonal channels provided by additional bandwidth. As discussed in \cite[section 4.2.2]{tse-2005}, there is no free lunch and a careful study of the network topology and shadowing conditions is needed to ensure noise-level multiuser interference.
Maintaining such orthogonality in the frequency domain may be hard for emergency notification application in URLLC. Furthermore, we argue that the energy of FDMA can be the same as that of TDMA if we keep the same usage of the total bandwidth. Assume that in both FDMA and TDMA the blocklengthes of user 1 and 2 are $m_1$ and $m_2$ while the longest user latencies are both $D$. For FDMA, besides $m_1 \leq D,  m_2 \leq D$, we also need $m_1+m_2 \leq D$ to keep the usage of the total bandwidth the same as that of TDMA. Then, the energy optimization problem of FDMA will be the same as that of TDMA. From Fig. 9, the proposed NOMA outperforms TDMA (and thus FDMA) under this setting.}. From the discussions on URLLC in 3GPP \cite{3GPP-URLLC1}\cite{3GPP-URLLC2}, we assume that the packets contain equal size of $32$ bytes. Also from \cite{3GPP-URLLC1}\cite{3GPP-URLLC2}, blocklengths $256$, $384$ and $640$ are adopted for QPSK modulations with channel code rates $1/2$, $1/3$ and $1/5$  respectively. These will be served as benchmarks to choose blocklengthes for users $D_1$ and $D_2$ in our following simulations. The block error probability of each user is set to be (around) $10^{-6}$. For NOMA and the hybrid scheme, we set $\epsilon_1 = \epsilon_2 = 10^{-6}$ and $\epsilon_{21} = \epsilon_{22} = 5\times10^{-7}$ such that the error probabilities in \eqref{eq_user2errSmallh1} and \eqref{eq_user2errHybrid} are both around $10^{-6}$.  We assume the channel coefficient is composed by the large-scale path loss and the small-scale Rayleigh fading. In particular, the distance-dependent path loss is modeled by $10^{-3} d^{-\alpha}$ where $d=10$ meter is the Euclidean distance between the transmitter and receiver and $\alpha=2$ is the path loss exponent; and the variance of the small-scale Rayleigh fading is unity. The energy is obtained by averaging 1000 channel realizations, if without specification. The system bandwidth is $1$ MHz and the noise power density is set to be  $\sigma_1^2 = \sigma_2^2 = -110$ dBm.
When $h_1>h_2$, both problems  \eqref{P-noma-case21} and \eqref{P-noma-case22} are solved and the one that yields the smaller energy  is chosen as the consumed energy of the NOMA scheme. The energy of TDMA is solved from the successive upper-bound minimization (SUM) method in \cite{Xu-2016}.

%


\begin{figure}[!tp]
	\centering
	\includegraphics[width=1.0\linewidth]{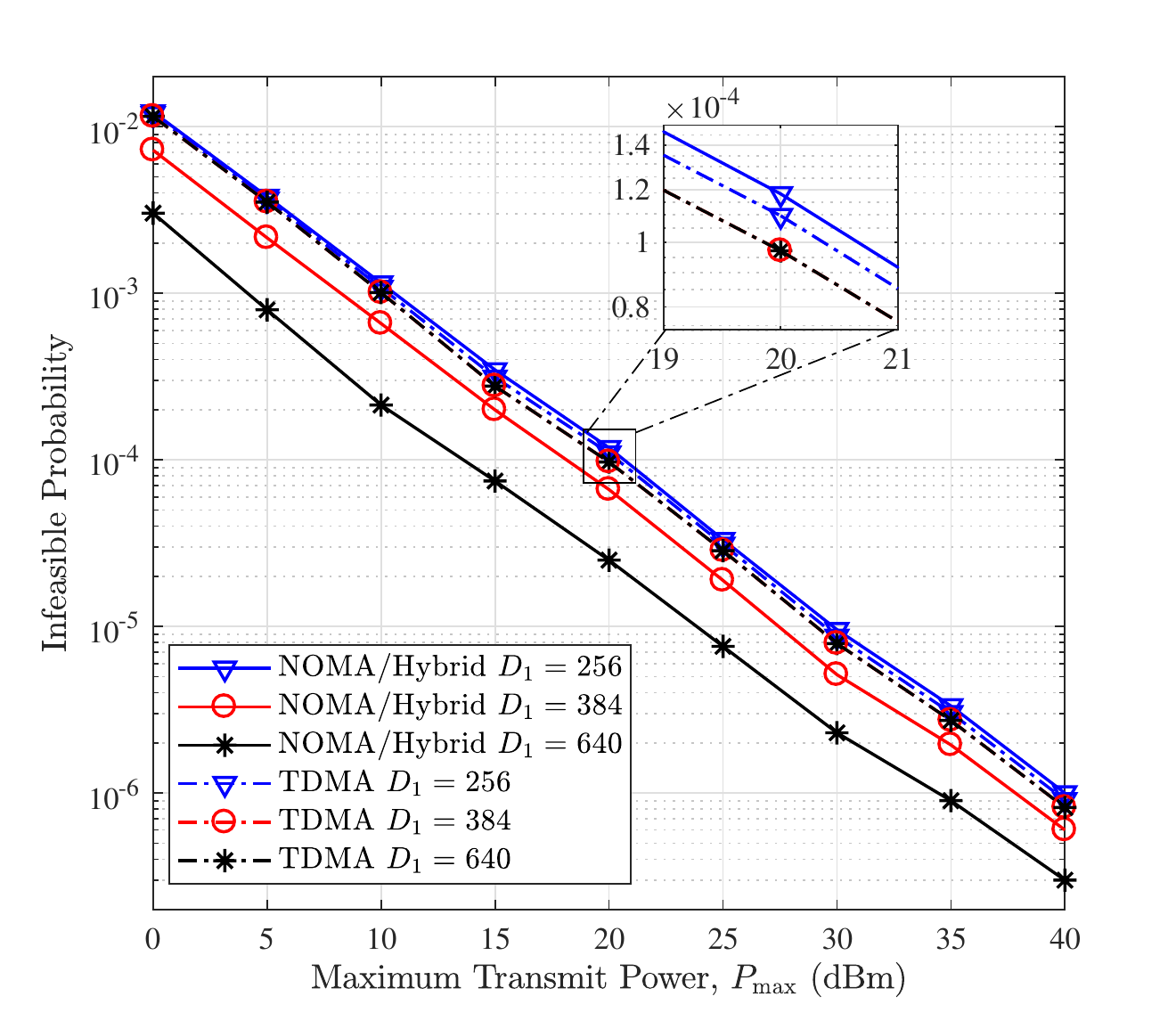}\\
	\caption{Infeasible probabilities of proposed transmission schemes with $D_2 = 640$.} \label{fig:Infeasible_Prob}
\end{figure}

As noted in Remark \ref{rem:reliability}, the reliability is determined by the feasibility probability of the optimization problem and the decoding error probability of each user. We set the communication reliability requirement of each user as $1 - 10^{-5}$. With $10^{-6}$ block error probability, the maximum infeasible probability of the optimization problem is approximately $9\times10^{-6}$. Now we evaluate the feasibility probability of each scheme and determine the corresponding $P_{\max}$.
Benefit from the feasibility conditions in Remark \ref{rem:feasibility}, the feasibility probabilities for NOMA and hybrid schemes are easy to find.
Remind that NOMA is just a special cases of the hybrid scheme, thus the feasibility of the later is also checked.
The feasibility of TDMA can be checked by checking that of the SUM solver in \cite{Xu-2016} while the corresponding complexity is much larger than Remark \ref{rem:feasibility}. Fig. \ref{fig:Infeasible_Prob} shows the probabilities when TDMA problem and NOMA problems \eqref{P-noma-case1}\eqref{P-noma-case21}\eqref{P-noma-case22} are infeasible, for different values of latency constraints and maximum available power $P_{\max}$ by performing $10^7$ channel realizations. Note that when $P_{\max} \ge 35$ dBm, the infeasible probabilities are all smaller than $4 \times10^{-6}$, thus the overall communication reliability can be satisfied. We thus chose $P_{\max} \ge 35$ dBm in the following simulations.
From Fig. \ref{fig:Infeasible_Prob}, one can observe that the infeasible probabilities decrease with the increase of $D_1$. It can also be observed that when $D_1 = 256$, the TDMA is a better option; while when $D_1 = 384$ and $640$, the NOMA schemes can outperform the TDMA due to that SIC can be efficiently performed at receivers. As pointed out by \cite{polar-tse} that, with only a 0.25dB SNR loss, the information-theoretic rate/capacity result in \cite{2010-Polyasiki-TIT} can be practically approached via the polar code. Thus to approximately evaluate the energy consumption of a real communication system, one can plus the SNR loss to the results under FBC\footnote{In practice, to implement the blocklength suggested by this paper, shortening or puncturing the polar code may be needed. However, the detailed design of such a code  is not trivial and beyond the scope of this work.}. For example,  with $D_1 = D_2 = 640$, the transmission energy computed using the above FBC capacity formula is $8.76$, where the energy is computed by the multiplication of the transmit power and the blocklength. Taking the SNR loss into consideration, the transmission energy using polar coding can be approximated to be $9.28$.

\begin{figure}[!tp]
	\subfigure[Energy consumption averaged for cases where $h_1 < h_2$.]{
		\includegraphics[width=0.5\textwidth]{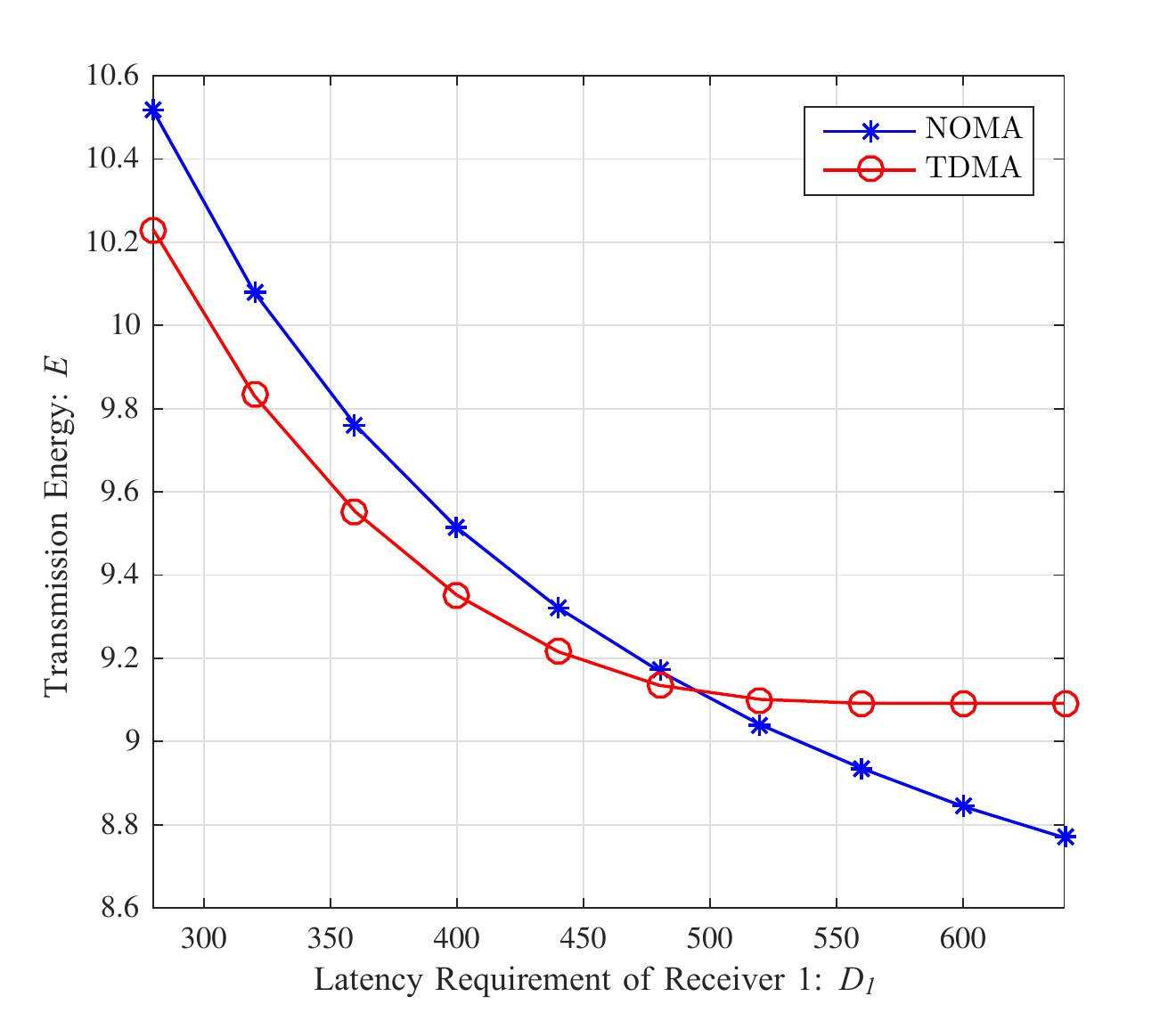}\label{fig:energy_comp_h2_larger}}
	\hspace{8mm}
	\subfigure[Energy consumption averaged for cases where $h_1 \ge h_2$.]{
		\includegraphics[width=0.5\textwidth]{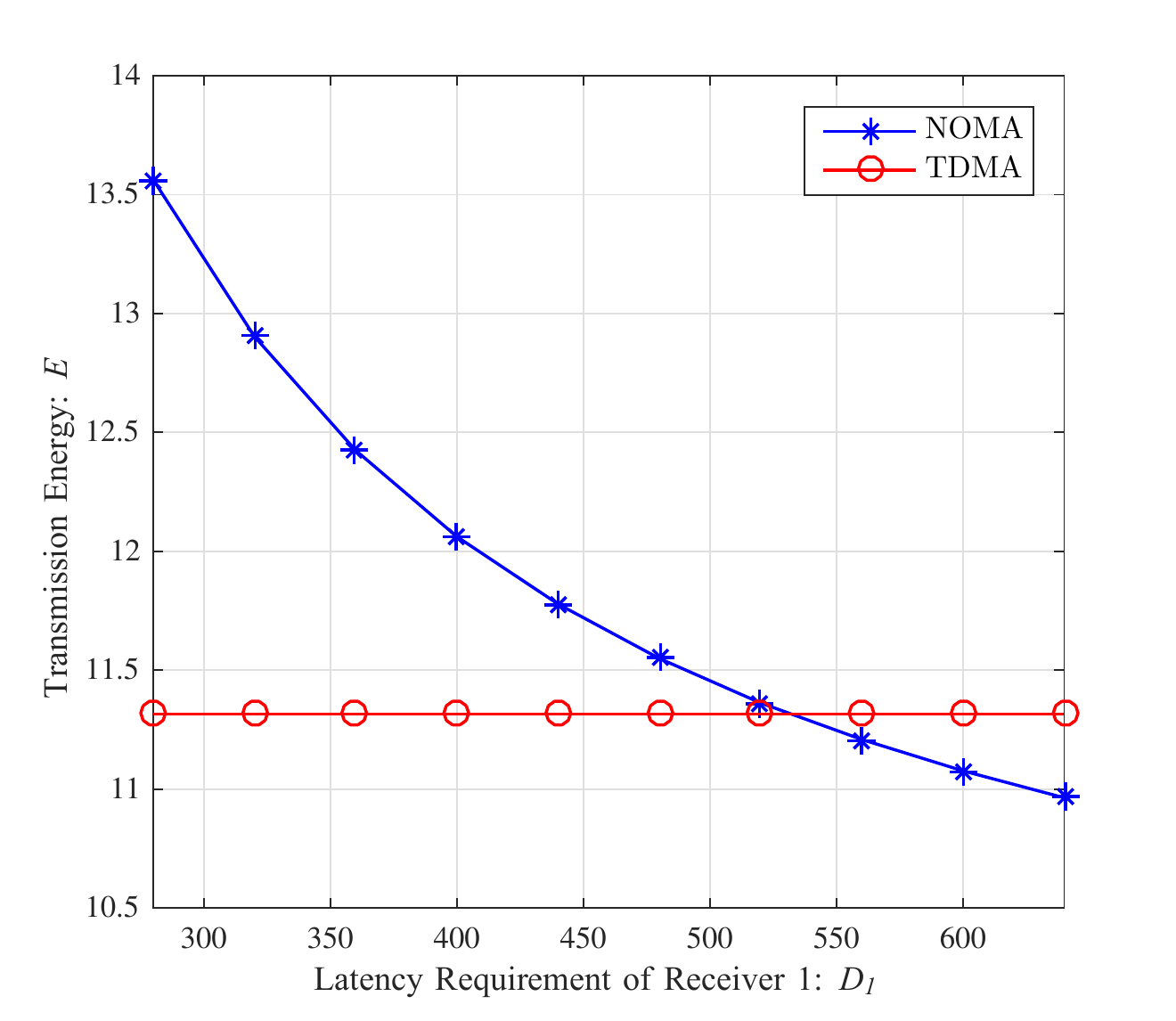} \label{fig:energy_comp_h1_larger}}\\
	\caption{Comparisons of consumed energy under NOMA and TDMA with $D_2 = 640$ and $P_{\max} = 40$ dBm.} \label{fig:energy_comp1}
\end{figure}
Fig. \ref{fig:energy_comp1} compares the transmission energy of the NOMA and TDMA schemes with different latency requirements of user 1, both enjoy low complexity resource allocation algorithms (if TDMA is feasible) without exhaustive linear searches. In Fig. \ref{fig:energy_comp_h2_larger}, we average the cases of $h_1 < h_2$ while in Fig. \ref{fig:energy_comp_h1_larger} we average the cases of $h_1 \ge h_2$. As it can be seen in Fig. \ref{fig:energy_comp_h2_larger}, the transmission energies for both schemes decline with the increase of $D_1$. Specifically, for the NOMA scheme, the transmission energy is strictly decreasing with $D_1$. The reason is that, from Section \ref{sec_NOMA} the optimal blocklength $m_1^*$ of user 1 is $D_1$, then a larger $D_1$ allows a longer $m^*_1$ and thus a smaller power and energy for delivering the packet for user 1. The consumed energy for user 2 is also decreased due to the reduced interference from user 1, and together resulting strictly decreasing energy with increasing $D_1$. While for the TDMA scheme, unlike NOMA, the optimal blocklength $m_1^*$ does not always equal to $D_1$. When $D_1$ is small, same as NOMA the optimal $m_1^* = D_1$, and the transmission energy decreases when $D_1$ increases. However, when $D_1$ is large enough, $m_1^* < D_1$ and $m_1^*$ becomes a constant when  $D_1$ increases. As a result, the transmission energy decreases first and then keeps unchanged in TDMA.
From Fig. \ref{fig:energy_comp_h1_larger}, we can observe that the consumed energy of the TDMA scheme is almost a constant when $D_1 \ge 280$, indicating that the schedule of user 1 is finished before $D_1$ with high probability due to the good channel condition $h_1$. Finally, it is important to note that when $D_1$ approaches $D_2$, the NOMA scheme gradually performs better than the TDMA scheme since SIC can effectively reduce the interference as long as it is feasible.

\begin{figure}[!tp]
	\centering
	\includegraphics[width=1.0\linewidth]{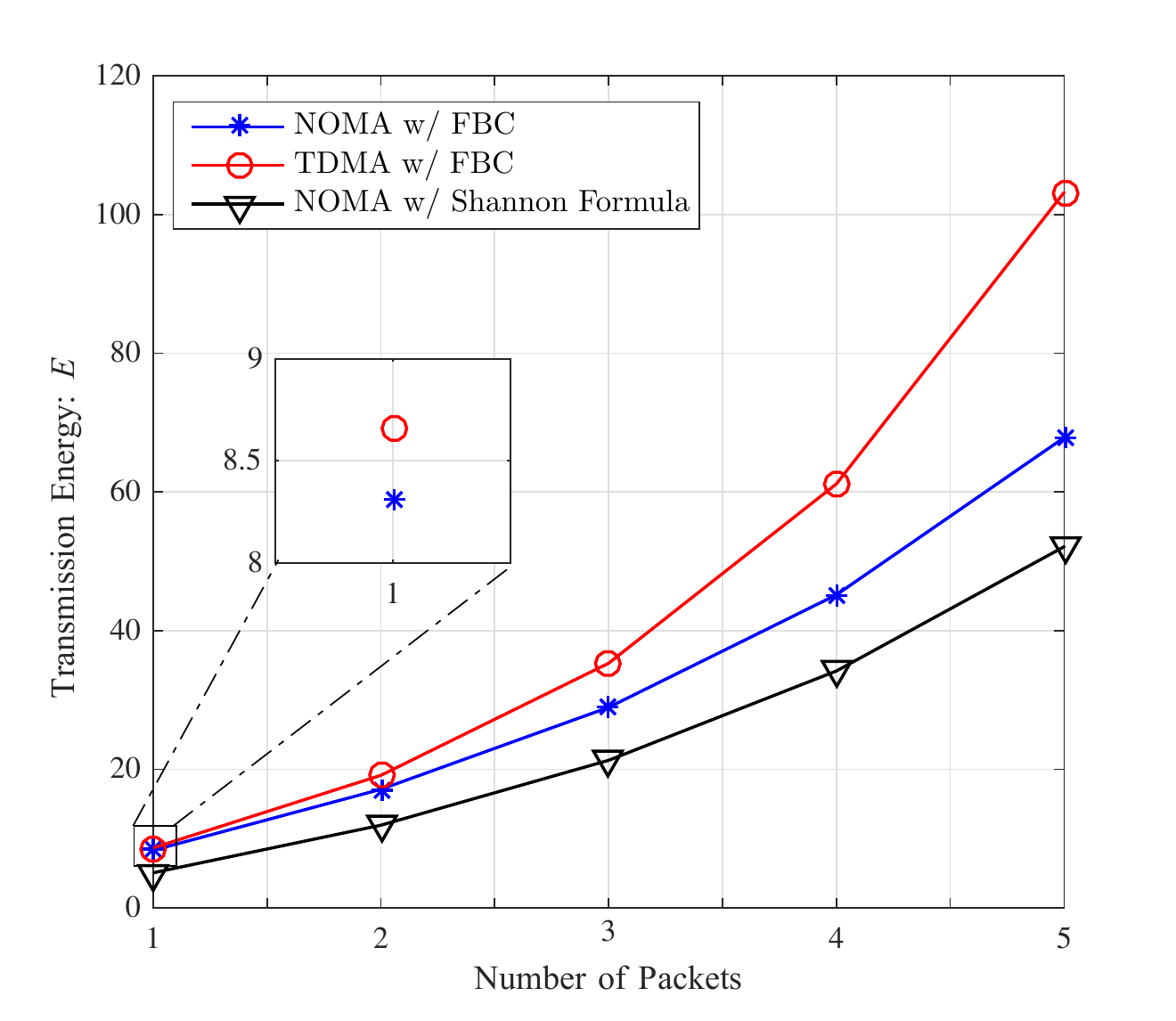}\\
	\caption{Energy consumption of various schemes using the normal approximation of  FBC and Shannon capacities versus the number of combined packets, with $D_1 = D_2 = 640$, $P_{\max} = 40$ dBm.} \label{fig:energy_vs_packet_number}
\end{figure}

\begin{figure}[!tp]
	\subfigure[Energy consumption with $D_2\!=\!640$ and $N\!=\!32$, by averaging $1000$ channel realizations.]
	{\includegraphics[width=0.5\textwidth]{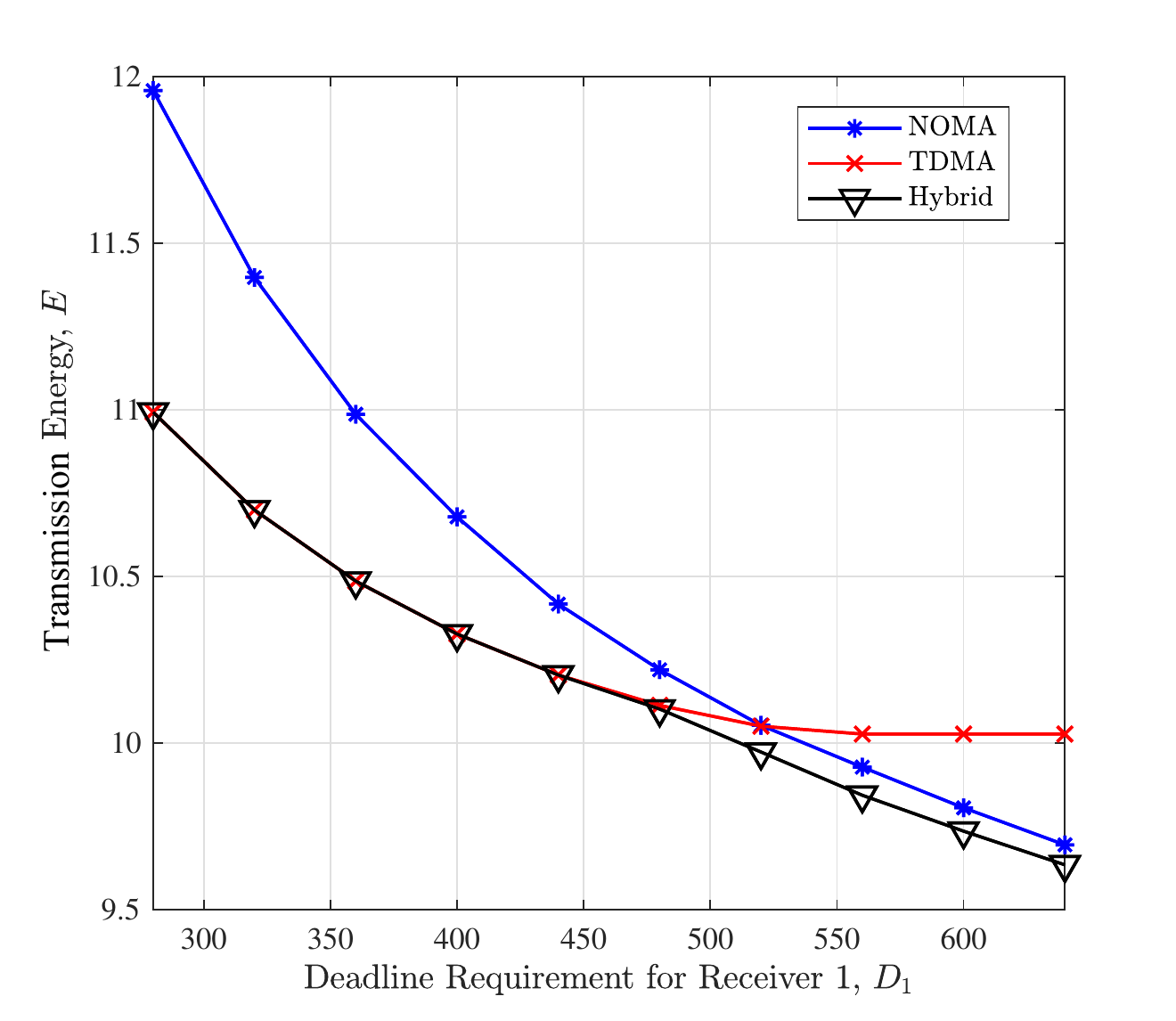}\label{fig:hybrid_energy1}}\\
	\hspace{8mm}
	\subfigure[Energy consumption with $D_2=640,$  $N=32*3$, and channel realization $h_1 = 300$ and $h_2 = 30$.]
	{\includegraphics[width=0.5\textwidth]{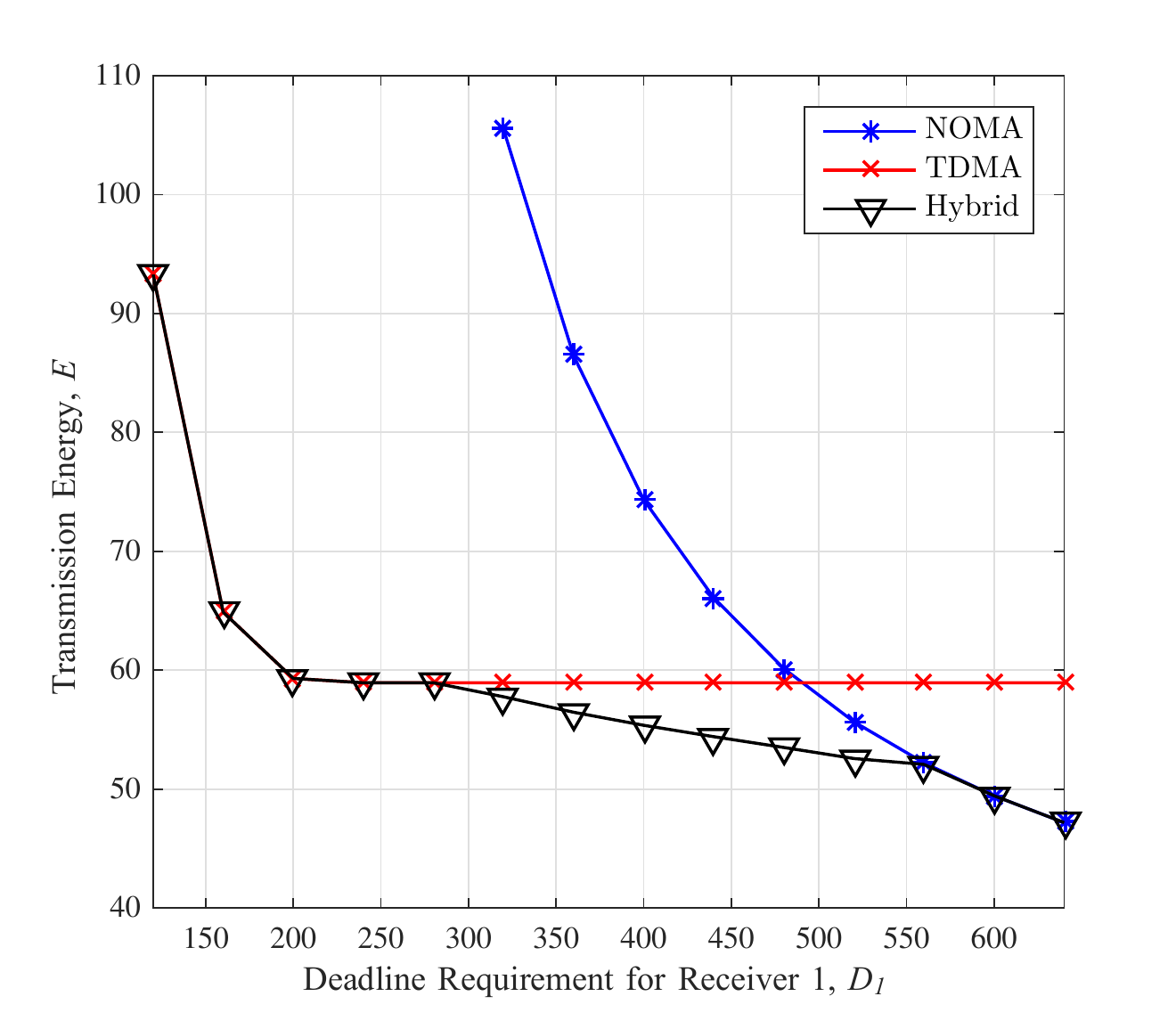} \label{fig:hybrid_energy2}}\\
	\caption{Energy consumption of proposed transmission schemes with $P_{\max} = 40$ dBm and $D_2 = 640$.} \label{fig:hybrid_energy}
\end{figure}

  We also compares the energy consumptions predicted by invoking the normal approximation of FBC capacity formula and Shannon capacity in Fig. \ref{fig:energy_vs_packet_number}, by fixing latency requirements but varying numbers of transmitted packets combined. 
Note that it was already shown that using Shannon capacity formula will underestimate the energy for TDMA \cite{Xu-2016}, we only show the energy consumption of NOMA by using the Shannon capacity. Again, using Shannon capacity still results in under-estimation of the energy consumption in NOMA. Furthermore, Shannon capacity may not predict the feasibility of SIC and success probabilities may increases. It can be also observed that the performance gain of NOMA compared to TDMA increases with the increase of the number of transmitted packets due to higher spectrum efficiency from non-orthogonal super-position coding. Finally, to investigate the energy consumption of the proposed hybrid scheme, in Fig. \ref{fig:hybrid_energy1}, it is compared with those from pure NOMA and TDMA when the packet size is $32$ bytes. One can observe that the proposed hybrid transmission scheme possesses the advantages of both the TDMA and NOMA scheme.
  Moreover, the hybrid scheme can be even more promising  when the packet size is lager than $32$ bytes. For example, in Fig. \ref{fig:hybrid_energy2}, when the combined packet size is $32\times3 = 96$ bytes,  the hybrid scheme can be ``strictly'' better than both NOMA and TDMA schemes even when $D_1$ is only half $D_2$, that is, $D_1=320$.
  The simulations for Fig 10(b) are performed under $P_{\max} = 40$ dBm, where the infeasible probability is smaller than $7 \times 10^{-6}$ when $D_1=256$.


\section{Conclusions} \label{sec_con}
In this paper, we have considered the energy-efficient transmission design problems subject to heterogeneous and strict latency and reliability constraints at receivers.
The normal approximation of FBC has been adopted to explicitly describe the trade-off between latency and reliability.
We first investigated the NOMA scheme.
However, due to the heterogeneous latency requirements, traditional SIC scheme may not be valid. To cope with this, novel interference mitigation schemes have been proposed.
Then by well utilizing the structure of the formulated nonconvex problems in NOMA schemes, optimal transmission powers and code block lengths have been derived, and the feasibility test and resource allocation are with low complexity. We have found that, due to the heterogeneous latency, pure NOMA scheme may not achieve the best transmission energy as the traditional homogeneous setting. In view of this, we have presented a hybrid scheme which can include the TDMA and NOMA as special cases. While the problem is difficult to solve, by approximating the normal approximation of FBC capacity formula, we have proposed the suboptimal but computationally efficient algorithm. The simulation results have shown that the hybrid scheme can possess the advantages of both NOMA and TDMA.

\setcounter{TempEqCnt}{\value{equation}} 
\setcounter{equation}{46} 
\begin{figure*}[!tp]  
	\begin{subequations}\label{first_derivative}
		\begin{align}
		\frac{d E}{d m} &\circeq x F_{x}^{\prime} - mF_m^{\prime} = \frac{mx}{x+1} - \frac{Q \sqrt{m}x}{(x+1)^2 \sqrt{x(x+2)}} - m\ln(x+1) + \frac{Q\sqrt{m}}{2} \frac{\sqrt{x(x+2)}}{x+1}\\
		&= \left(\frac{x}{x+1} -\ln(x+1)\right)m + \frac{(x+1)(x(x+2)) - 2x}{2(x+1)^2\sqrt{x(x+2)}}Q\sqrt{m}\\
		&\circeq \left(\frac{x}{x+1} -\ln(x+1)\right){\sqrt{m}} + \frac{x^3+3x^2}{2(x+1)^2\sqrt{x(x+2)}}Q   \label{deriv_1}\\
		&= \left(\frac{x}{x+1} -\ln(x+1)\right)\frac{\sqrt{x(x+2)}Q + c}{2(x+1)\ln(x+1)} + \frac{x^3+3x^2}{2(x+1)^2\sqrt{x(x+2)}}Q \label{deriv_0}\\
		&= \frac{x\sqrt{x(x+2)} Q + cx}{2(x+1)^2 \ln(x+1)} - \frac{\ln(x+1)\sqrt{x(x+2)}Q + c\ln(x+1)}{2(x+1) \ln(x+1)} + \frac{x^3+3x^2}{2(x+1)^2\sqrt{x(x+2)}} Q\\
		&\circeq x^2(x\!+\!2)Q \!+\! cx\sqrt{x(x\!+\!2)}\!-\! x(x\!+\!1)(x\!+\!2)\ln(x\!+\!1)Q - c(x\!+\!1)\sqrt{x(x\!+\!2)}\ln(x\!+\!1)+ (x^3+3x^2)\ln(x+1)Q \label{deriv_2}\\
		&= \big(x^2(x\!+\!2)\!-\!(x^3\!+\!3x^2\!+\!2x)\ln(x\!+\!1)\!+\!(x^3\!+\!3x^2)\ln(x\!+\!1) \big)Q \!+\! \big(x\!-\!(x\!+\!1)\ln(x\!+\!1)\big)\sqrt{x(x\!+\!2)}c\\
		&= \big(x^2(x\!+\!2) \!-\! 2x\ln(x\!+\!1)\big) Q \!+\!\! \big(x\!-\!(x\!+\!1)\ln(x\!+\!1)\big)\!\sqrt{x(x\!+\!2)}\!\sqrt{x(x\!+\!2)Q^2 \!+\! 4N \!(x\!+\!1)^2 \!\ln(x\!+\!1)\!\ln 2}\\
		&\le \big(x^2(x\!+\!2)\! -\! 2x\ln(x\!+\!1)\big) Q \!+\! \frac{\sqrt{2}}{2}\big(x\!-\!(x\!+\!1)\ln(x\!+\!1)\big)\sqrt{x(x\!+\!2)}\left(\!\!\sqrt{x(x\!+\!2)Q^2} + \sqrt{4N (x\!+\!1)^2 \ln(x\!+\!1)\ln 2}\right) \label{deriv_3}\\
		&\le \left(\!\!x^2(x\!+\!2) \!-\! 2x\frac{2x}{x\!+\!2}\!\!\right)\!Q  \!+\!\frac{\sqrt{2}}{2}\!\!\left(\!\!x\!-\!(x\!+\!1)\frac{2x}{x\!+\!2}\!\!\right)\!x(x\!+\!2)Q \!+ \!\sqrt{2}\!\left(\!x\!-\!(x\!+\!1)\frac{2x}{x\!+\!2}\!\right)\!(x\!+\!1)\! \sqrt{x(x\!+\!2)} \sqrt{\frac{2x}{x\!+\!2}N\ln 2} \label{deriv_4}\\
		&= \left(x^2(x+2)^2 - 4x^2\right)(x+2) Q - \frac{\sqrt{2}}{2}x^3(x+2)^2Q - \sqrt{2}x^2(x+1) \sqrt{x(x+2)} \sqrt{2x(x+2)N\ln 2}\\
		&= x^3(x+4)(x+2) Q - \frac{\sqrt{2}}{2}x^3(x+2)^2Q - 2 x^3(x+1) (x+2) \sqrt{N\ln 2}\\
		&\circeq 2(x+4) Q - \sqrt{2}(x+2)Q - 4(x+1) \sqrt{N\ln 2}   \label{deriv_5}\\
		&= \underbrace{(2Q-\sqrt{2}Q - 4\sqrt{N\ln 2})}_{f_1(Q,N)}x + \underbrace{8Q-2\sqrt{2} Q -4 \sqrt{N\ln 2}}_{f_2(Q,N)}.
		\end{align}
	\end{subequations}
	\hrulefill
\setcounter{equation}{43} 
\end{figure*}

\begin{appendices}

	\section{Proof of Lemma \ref{lem:energy_monotonicity} }\label{app1}
	For the simplicity of notations, we remove the subindex of all variables and let $x$ denote the SINR.
	Based on \eqref{p1.1}, we define
	\begin{align}\label{func_F}
    & F(m,x) \triangleq \nonumber \\
	& ~~m\ln(x + 1) - \sqrt{m} \frac{\sqrt{x(x+2)}}{(x + 1)} Q^{-1}(\epsilon) - N \ln2 = 0,
	\end{align}
	and the partial derivatives of $F(m,x)$ with $m$ and $x$ can be described as
	\begin{subequations}
		\begin{align}
		F_m^{\prime} &= \ln(x+1) - \frac{Q}{2\sqrt{m}} \frac{\sqrt{x(x+2)}}{x+1},\\
		F_{x}^{\prime} &= \frac{m}{x+1} - \frac{Q \sqrt{m}}{(x+1)^2 \sqrt{x(x+2)}}.
		\end{align}
	\end{subequations}
	As is shown in \cite{Xu-2016}, $F_m^{\prime}>0$ and $F_{x}^{\prime}>0$ always hold  with $m>0$ and $x>0$ respectively. The monotonicity of $E(m) = m \Gamma(m)$, where $\Gamma(m)=x$,  can be verified by checking the sign of its first derivative. From the implicit function theorem \cite{Krantz_Parks02}, we have
	\begin{align}
	\frac{d E}{d m}  = \frac{d mx}{d m} \!=\! x \!+\! m\frac{d x}{d m} \!=\! x \!-\! m\frac{F_m^{\prime}}{F_{x}^{\prime}} \!\circeq\! x F_{x}^{\prime} - mF_m^{\prime},
	\end{align}
	where $A \circeq B$ denotes that $A$ and $B$ have the same sign. The sign of $\frac{d E}{d m}$ is checked in \eqref{first_derivative}. Note that the right-hand side of \eqref{func_F} is a quadratic equation of $\sqrt{m}$, and by letting $Q = Q^{-1}(\epsilon)$ and $c = \sqrt{x(x+2)Q^2 + 4(x+1)^2 \ln(x+1) N \ln 2 }$, the positive root $\sqrt{m}$ given $x$ in \eqref{deriv_1} is
	\setcounter{equation}{47} 
	\begin{align}
	\sqrt{m} = \frac{\sqrt{x(x+2)}Q + c}{2(x+1)\ln(x+1)}
	\end{align}
	and it results in \eqref{deriv_0}; also \eqref{deriv_1} and \eqref{deriv_2} hold due to $\sqrt{m} >0$ and $2\sqrt{x(x+2)}(x+1)^2\ln(x+1)>0$ for $x>0$ respectively; \eqref{deriv_3} holds because of $x-(x+1)\ln(x+1)<0$ with $x>0$ and the fact that $2\sqrt{a+b} \ge \sqrt{2}\big(\sqrt{a}+\sqrt{b}\big)$ for $a,b>0$; \eqref{deriv_4} holds owing to $\ln(x+1)\ge \frac{2x}{x+2}$ for $x>0$. In addition, \eqref{deriv_5} holds since $x^3(x+2) > 0$ for $x>0$.
	
	To prove that ${E}(m)$ is a monotonically decreasing function, we need $f_1(Q,N)<0$ and $f_2(Q,N)<0$ in \eqref{first_derivative}, indicating
	\begin{equation}\label{eq_mono_condition_App}
	\frac{Q}{\sqrt{N}} \le \frac{2\sqrt{\ln2}}{4-\sqrt{2}}
	\end{equation}
	Note that both $f_1(Q,N)$ and $f_2(Q,N)$ increase with $Q$ and decrease with $N$, and $Q = Q^{-1}(\epsilon)$ is a monotonically decreasing function with $\epsilon$. Therefore, for the pair $(\epsilon,N)$ satisfying \eqref{eq_mono_condition_App}, if we increase $\epsilon$ and $N$, the monotonicity of ${E}(m)$ also holds. This completes the proof.
	\hfill $\blacksquare$

	\section{Proof of Theorem \ref{thm:noma_case1}} \label{app:thm1}
We first claim that \eqref{p1.3} must hold with equality at the optimum, i.e., the optimal blocklengths must be
$m_k^* = D_k$ for all $k=1,2.$	Suppose that this is not true, i.e., $m_k^* < D_k$ for $k=1$ or $k=2$. Then one can further increase $m_k^*$.
According to  \cite[Proposition 1]{Xu-2016}, $\Gamma_k(m_k)$ is monotonically decreasing with $m_k > 0$.
Since
\begin{equation}\label{eq_p1p2_feasible_smallh1}
p_1+p_2=\frac{\Gamma_1(m_1) \Gamma_2(m_2)}{h_2}+\frac{\Gamma_1(m_1)}{h_1}+\frac{ \Gamma_2(m_2)}{h_2},
\end{equation}
$p_1^*+p_2^*$ can be reduced without violating \eqref{p1.4} when $m_k^*$ increases.
Besides, by Lemma \ref{lem:energy_monotonicity}, we also know that $m_k\Gamma_k(m_k)$ is decreasing with $m_k$. Thus the energy function in \eqref{eq_P_MOMA_target} can be reduced when $m_k^*$ increases. These two facts  contradict with the optimality of $m_k^*$. So we must have $m_k^* = D_k$ for all $k=1,2.$ Correspondingly, $\gamma_k^*= \Gamma_k(m^*_k)$ from \eqref{eq_Gamma_P_MOMA} and $p_k^*$ can be obtained from \eqref{eq:p_gamma} accordingly, which lead to  the optimal solution in \eqref{optimal_solution_case1}.

Finally, note that $\gamma_k= \Gamma_k(m_k)$ from \eqref{eq_Gamma_P_MOMA} is strictly decreasing with $m_k$ \cite[Proposition 1]{Xu-2016}, and thus the optimal and unique SINR $\gamma_k^*= \Gamma_k(D_k)$ can be efficiently computed
by the bisection search in Algorithm \ref{alg:bisection}
\cite{Boyd-2009}. 	\hfill $\blacksquare$

\section{Proof of Proposition \ref{prop:convexity_FBC} }\label{app2}
	Due to $f(x) \ge 0$, it needs
	\begin{align}
	a \le \frac{(x+1)\ln(x+1)}{\sqrt{x(x+2)}} \triangleq g(x)
	\end{align}
	To verify the monotonicity and convexity of $f(x)$, we give its first and second-order derivatives as follows 
	\begin{subequations}
		\begin{align}
		f'(x) &= \frac{1}{x+1} -\frac{a}{(x+1)^2 \sqrt{x(x+2)}},\\
		f''(x) &= \frac{-1}{(x+1)^2} + \frac{a(2(x(x+2)) + (x+1)^2)}{(x+1)^3 (x(x+2))^{\frac{3}{2}}}
		\end{align}
	\end{subequations}
	To guarantee $f(x)$ a monotonically increasing function, we require $f'(x) \ge 0$. Thus it needs
	\begin{align}
	a \le (x+1)\sqrt{x(x+2)} \triangleq g_1(x)
	\end{align}
	It can be easily proved that $g_1(x) \ge g(x)$ for $x\ge 0$, which implies that if the finite blocklength capacity formula holds then $f(x)$ is a monotonically increasing function. Further, to guarantee $f(x)$ a concave function, we require $f''(x) \le 0$ and have
	\begin{subequations}
		\begin{align}
		f''(x)
		&= \frac{-1}{(x+1)^2} + \frac{a(2(x(x+2)) + (x+1)^2)}{(x+1)^3 (x(x+2))^{\frac{3}{2}}} \\
		&= \frac{a(3x^2+6x+1) - (x+1)(x(x+2))^{\frac{3}{2}}}{(x+1)^3 (x(x+2))^{\frac{3}{2}}} \\
		&\circeq a(3x^2+6x+1) - (x+1)(x(x+2))^{\frac{3}{2}} \le 0
		\end{align}
	\end{subequations}
	or equivalently
	\begin{align}
	a \le \frac{(x+1)(x(x+2))^{\frac{3}{2}}}{3x^2+6x+1} \triangleq g_2(x)
	\end{align}
	On the contrary, when $a \ge g_2(x)$, $f(x)$ is convex.
	
	With some algebraic manipulations, we can find that $g(x)$ and $g_2(x)$ are monotonically increasing functions, and $g_2(x) \le g(x)$ for $0 \le x \le x_0$; $g_2(x) > g(x)$ for $x>x_0$ where $x_0 = 0.6904$ is the positive solution of equation $g_2(x) = g(x)$.
	Therefore, for given parameter $a$ and defining $\beta \triangleq g(x_0)= g_2(x_0)$, if $a>\beta$, $f(x)$ is concave for $x\ge g^{-1}(a)$; if $a \leq \beta$, $f(x)$ is convex for $g^{-1}(a) \le x \le g_2^{-1}(a)$ and concave for $x>g_2^{-1}(a)$. This completes the proof.
	\hfill $\blacksquare$
		
\end{appendices}

{
	\smaller[1]

}

\end{document}